\documentclass[aip,graphicx]{revtex4-1}

\usepackage{amsmath}
\usepackage{amsfonts}
\usepackage{amssymb}
\usepackage{graphicx}
\usepackage{epstopdf}
\usepackage{float}
\usepackage{color}

\draft % marks overfull lines with a black rule on the right

\begin{document}

%\preprint{}

\title{Electrostatic and whistler instabilities excited by an electron beam} %Title of paper



\author{Xin An}
\email[]{xinan@atmos.ucla.edu}
%\thanks{}
\affiliation{Department of atmospheric and oceanic sciences, University of California, Los Angeles}

\author{Jacob Bortnik}
\affiliation{Department of atmospheric and oceanic sciences, University of California, Los Angeles}

%\author{George Morales}
%\affiliation{Department of physics and astronomy, University of California, Los Angeles}

\author{Bart Van Compernolle}
\affiliation{Department of physics and astronomy, University of California, Los Angeles}

\author{Viktor Decyk}
\affiliation{Department of physics and astronomy, University of California, Los Angeles}

\author{Richard Thorne}
\affiliation{Department of atmospheric and oceanic sciences, University of California, Los Angeles}


\date{\today}

\begin{abstract}
The electron beam-plasma system is ubiquitous in the space plasma environment. Here, using a Darwin particle-in-cell method, the excitation of electrostatic and whistler instabilities by a gyrating electron beam is studied in support of recent laboratory experiments. It is assumed that the total plasma frequency $\omega_{pe}$ is larger than the electron cyclotron frequency $\Omega_e$. The fast-growing electrostatic beam-mode waves saturate in a few plasma oscillations by slowing down and relaxing the electron beam parallel to the background magnetic field. Upon their saturation, the finite amplitude electrostatic beam-mode waves can resonate with the tail of the background thermal electrons and accelerate them to the beam parallel velocity. The slower-growing whistler waves are excited in primarily two resonance modes: (a) through Landau resonance due to the inverted slope of the beam electrons in the parallel velocity; (b) through cyclotron resonance by scattering electrons to both lower pitch angles and smaller energies. It is demonstrated that, for a field-aligned beam, the whistler instability can be suppressed by the electrostatic instability due to a faster energy transfer rate between beam electrons and the electrostatic waves. Such a competition of growth between whistler and electrostatic waves depends on the ratio of $\omega_{pe}/\Omega_e$. In terms of wave propagation, beam-generated electrostatic waves are confined to the beam region whereas beam-generated whistler waves transport energy away from the beam.
\end{abstract}


\maketitle %\maketitle must follow title, authors, abstract and \pacs

\section{Introduction}
Energetic electron beams are ubiquitous throughout the solar system, such as the upstream from the interplanetary shock \citep{Tokar1984-JGR, Marsch1985-JGR, Bale1999-GRL}, the auroral ionosphere \citep{Carlson1998-GRL, Maggs1978-JGR}, solar flares \citep{petrosian1973impulsive}, in the outflow region of magnetic reconnection \citep{drake2003formation, Pritchett2004-JGR} and possibly the Earth's outer radiation belt \citep{Li2016-GRL}. The electron beam provides a free energy source for generating various electrostatic and electromagnetic instabilities. For example, a finite amplitude single electrostatic wave can be excited by a small cold beam \citep{Oneil1971-PoF, Gentle1973-PoF}. Whistler waves can also be excited by an electron beam in a number of space plasma settings \citep{Maggs1976-JGR, Gary1977-JGR, Tokar1984-JGR, Huang2016-JGR}. Some electrostatic structures, such as double layers and electron holes, seems to be generated by current-carrying electron beams in the presence of density inhomogeneities \citep{Newman2001-PRL}. Artificial electron beams have been injected into the Earth's ionosphere and magnetosphere to probe the space environment and to study the rich variety of waves in the beam-plasma interaction (see Ref.~\onlinecite{Winckler1980-ROG} and references therein). Extensive laboratory experiments in the past have been conducted to study the beam-generated whistler waves \citep{Stenzel1977-JGR, Krafft1994-PRL, Staro1999-PRL} and electrostatic waves \citep{Gentle1973-PoF}. Accordingly, many numerical experiments utilizing the particle-in-cell method were devoted to study the wave instabilities excited in the electron beam-plasma interaction \citep{Morese1969-PoF, Omura1987-JGR, Omura1988-GRL, Pritchett1989-GRL, Gary2000-PoP, Fu2014-PoP, Che2017-PNAS}.\\

A series of controlled laboratory experiments \citep{VanCompernolle2015-PRL, *VanCompernolle2016-PRLerr, An2016-GRL, VanCompernolle2017-PPCF} were performed to study the excitation of whistler waves in the Large Plasma Device \citep{Gekelman2016-RSI} at University of California, Los Angeles (UCLA). In the experiments, both electrostatic and whistler waves were excited by the injection of a gyrating electron beam into a cold plasma. It was demonstrated that the whistler mode waves were excited through a combination of cyclotron resonance, Landau resonance and anomalous cyclotron resonance \citep{An2016-GRL}. A measurement of the electron distribution function is desired to study the self-consistent wave-particle interactions. But such a diagnostic of the electron distribution is not available at the present time. On the other hand, linear kinetic theory can predict the growth rate of electrostatic beam-mode and whistler waves for a given beam distribution. But the linear theory cannot resolve how the linearly unstable waves modify the electron distribution and therefore cannot resolve the saturation of the beam instability. Moreover, since both electrostatic beam-mode and whistler waves can extract energy from the inverted slope ($\partial f_b / \partial v_\parallel > 0$, $f_b$ is the beam distribution function, $v_\parallel$ is the parallel velocity) of the electron beam through Landau resonance, the fast-growing of electrostatic beam-mode waves can affect the slow-growing whistler instabilities via this inverted population. Here, using a self-consistent Darwin particle-in-cell method, we study the excitation of electrostatic and whistler waves in a beam-plasma system, the associated evolution of the electron distribution and the competing growth between electrostatic beam-mode and whistler waves.

\section{Computational setup}
The Darwin particle-in-cell (PIC) model used in this study is based on a two-dimensional spectral code developed as part of the UCLA particle-in-cell (UPIC) framework \citep{Decyk2007-CPC, PICKSC-skeleton}. The Darwin PIC model has been used previously to study the whistler anisotropy instability in the solar wind \citep{Hughes2016-PoP} and Earth's inner magnetosphere \citep{Schriver2010-JGR, An2017-JGR}. Compared to a conventional electromagnetic PIC method, the Darwin PIC method excludes the transverse component of the displacement current in Ampere's law and hence excludes retardation effects and light waves, but leaves the physics of whistler waves unaffected \citep{BUSNARDONETO1977300, GEARY1986313, Hewett1985}. Thus the Darwin PIC model does not have the restriction on the time step set by Courant condition $\Delta t < \delta / c$. Here $\Delta t$ is the time step used in the simulation, $\delta$ is the grid spacing and $c$ is the speed of light. The grid spacing $\delta$ is required to resolve the Debye length to prevent numerical heating. Consequently, for a plasma having a background thermal component ($v_{t}/c = 0.01$) as in this study, the fully electromagnetic PIC method requires a very small time step ($\Delta t \lesssim 0.01\, \omega_{pe}^{-1}$) whereas the Darwin PIC method does not. Such an advantage greatly improves the computation efficiency.\\

A beam-plasma system with two dimensions of configuration space and three dimensions of velocity space is explored. The boundary conditions for both particles and fields are periodic in two spatial directions. The computational domain consists of $L_x = 4096$ grids in $x$ direction and $L_y = 1024$ grids in $y$ direction with a grid spacing of $0.02\, d_e$. Here $d_e = c/\omega_{pe}$ is the electron inertial length. $\omega_{pe}$ is the plasma frequency. Each cell contains $64$ particles, which is sufficient to keep a low level of particle noise and converge the growth rate of instabilities. The time step is $0.1\, \omega_{pe}^{-1}$. The total simulation time is $500\, \omega_{pe}^{-1}$ to include both the linear and nonlinear stages of the instabilities. A uniform external magnetic field $B_0$ is applied in the $x$ direction with a magnitude $\Omega_e/\omega_{pe} = 0.2$. In this study, the ions are immobile and form a charge neutralizing background. A beam ring distribution is initialized in the system, which takes the form
\begin{eqnarray}
f_b \propto e^{ - \frac{ \left( v_{\parallel} - V_{\parallel b} \right)^2 }{ 2 v_{tb\parallel}^2 } } e^{ - \frac{ \left( v_{\perp} - V_{\perp b} \right)^2 }{ 2 v_{tb\perp}^2 } }
\end{eqnarray}
It has a streaming velocity $V_{\parallel b}/c = 0.0766$ parallel to the magnetic field and a velocity ring centered at $V_{\perp b}/c = 0.0766$ in the perpendicular direction, which corresponds to an electron beam of $3$\,keV in kinetic energy and $45$ degree in pitch angle, which is typical in the experiment. The thermal spread of the beam is chosen as $v_{tb\parallel} = v_{tb\perp} = 0.001 c$ so that the beam has a narrow ``ring'' in both parallel and perpendicular directions, mimicking that of the experiment. The beam density profile is localized in the $y$ direction and uniform in the $x$ direction, which takes the form
\begin{eqnarray}\label{eqn:beam-density-profile}
n_b(y) = \begin{cases}
n_b, & \frac{3}{8}L_y<y<\frac{5}{8}L_y \\
0, & \mbox{otherwise}
\end{cases}
\end{eqnarray}
The beam width $L_y/4$ is about $13$ times larger than the gyro-radius of the beam electrons, which is comparable to that in the experiment. In the beam region, the ratio of the beam density $n_b$ to the total plasma density is $n_b/(n_b + n_0) = 1/8$, where $n_0$ is background plasma density in the beam region. Note that the ratio of beam density to total plasma density is about $0.001 \sim 0.005$ in the experiment, which is much lower than that in the simulation. Correspondingly, relevant quantities in the simulation, such as the linear growth rate and the saturation time of the waves, should be properly scaled to compare with that in the experiment. The background electrons form a return current that cancels the beam current in the parallel direction, i.e., $n_b V_{\parallel b} + n_0 V_{\parallel 0} = 0$. Here $V_{\parallel 0} = -V_{\parallel b}/7$ is the streaming velocity of background electrons in the beam region. Aside from this small streaming velocity in the beam region, the background electrons have an isotropic Maxwellian distribution with a thermal velocity of $0.01 c$ (about $50$\,eV in thermal temperature). Outside the beam region, the density of background electrons is $n_b + n_0$ so that the total plasma density is uniform.

\section{The wave field}
\begin{figure}[tphb]
\centering
\includegraphics[width=0.7\textwidth]{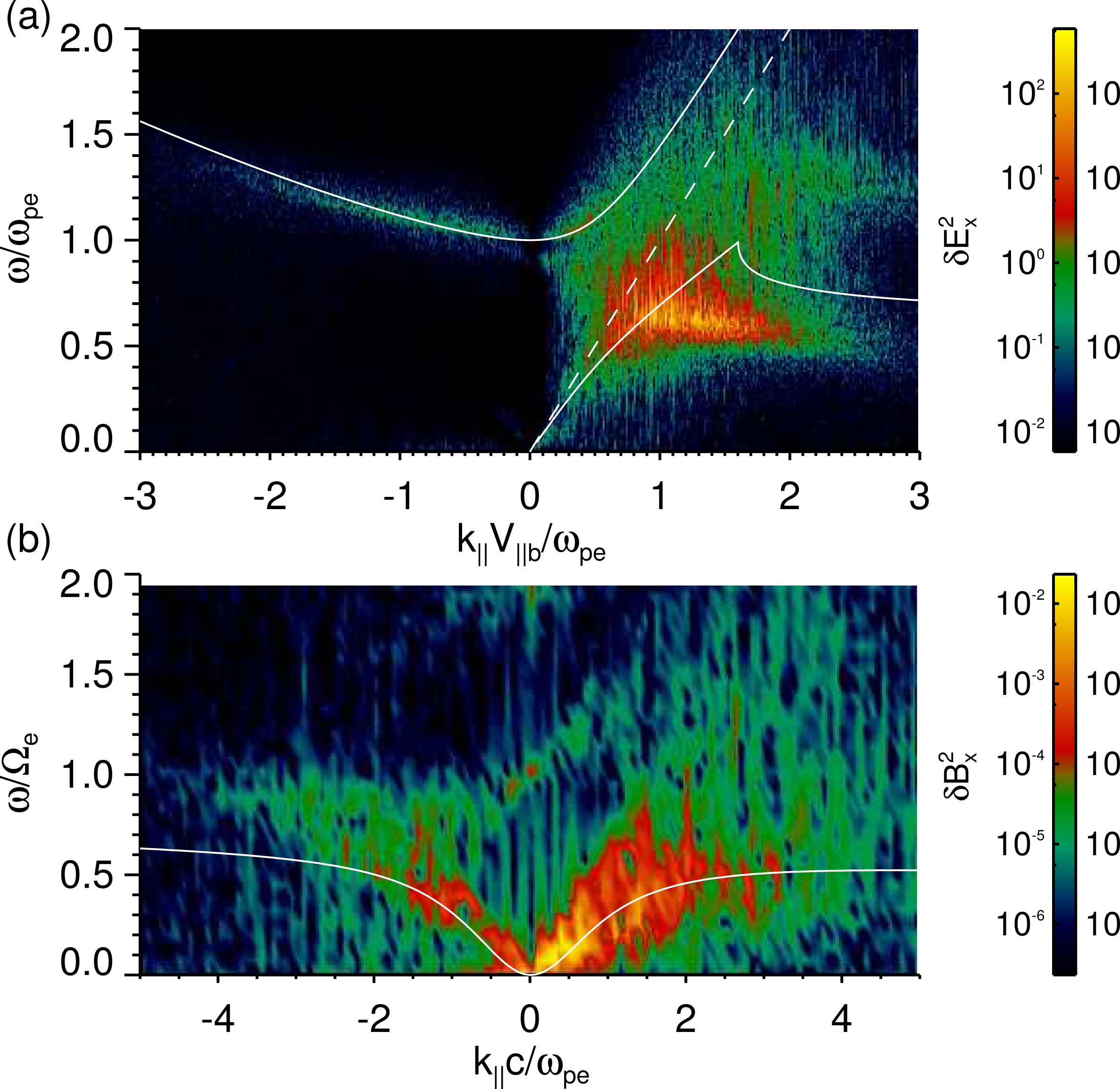}
\caption{(a) The power spectral density of $\delta E_x$. The solid white lines stand for the dispersion relation of electrostatic waves ($k_\perp = 0$) by solving equation \eqref{eqn-es-disp-rel}. The dashed white line represents $\omega = k_\parallel V_{\parallel b}$. (b) The power spectral density of $\delta B_x$. The solid white line stands for the dispersion relation of a whistler wave propagating at $55^\circ$ with respect to the background magnetic field in a cold plasma. Note that the parallel wave number $k_\parallel$ is normalized by $V_{\parallel b}/\omega_{pe}$ in panel (a) whereas it is normalized by $c/\omega_{pe}$ in panel (b), in order to better manifest the typical wave length of each wave.}
\label{fig5-dispersion-relation}
\end{figure}

A slice of wave field data, electric field $\delta E_x$ and magnetic field $\delta B_x$, is taken along the $x$ direction located at $y = L_y/2$ at every time step. The wave field $\delta E_x$ and $\delta B_x$ are Fourier-transformed to the space of $\omega$ - $k_\parallel$, where $\omega$ is the wave frequency and $k_\parallel$ is the parallel wave number. The power spectral density of $\delta E_x$ is shown in Figure \ref{fig5-dispersion-relation}a. Note that the magnetic power spectral density at the high frequencies around $\omega_{pe}$ is much weaker than the electric power spectral density. Thus the wave modes in Figure \ref{fig5-dispersion-relation}a are dominantly electrostatic. To identify the wave modes, the electrostatic dispersion relation is written as (assuming $k_\perp = 0$)
\begin{eqnarray}\label{eqn-es-disp-rel}
1 - \frac{\omega_{pe0}^2}{(\omega - k_\parallel V_{\parallel 0})^2} \left( 1 + 3 k_\parallel^2 \lambda_D^2 \right) - \frac{\omega_{pb}^2}{(\omega - k_\parallel V_{\parallel b})^2} = 0
\end{eqnarray}
where $\lambda_D$ is the Debye length of the thermal core electrons, $\omega_{pe0}$ is the plasma frequency of core electrons and $\omega_{pb}$ is the beam plasma frequency. Here the wave propagation is assumed to be parallel, i.e. $k_\perp = 0$, since the propagation angle is found to be within $20$ degrees with respect to the background magnetic field. For a given $k_\parallel$, the dispersion relation is solved for a complex wave frequency $\omega$. The real part of $\omega$ is shown for a spectrum of $k_\parallel$ as the white solid lines in Figure \ref{fig5-dispersion-relation}a. It is seen that the beam mode intersects with the Langmuir waves and modifies the topology of the dispersion relation of Langmuir waves. The electrostatic beam-mode waves has an enhanced power spectral density at $k_\parallel = 0.5 \sim 2\, \omega_{pe}/V_{\parallel b}$, which is consistent with the unstable range of the imaginary part of $\omega$ (not shown). Note that the intense electrostatic waves below $\omega_{pe}$ would not be present without an electron beam. The power spectral density of $\delta B_x$ is shown in Figure \ref{fig5-dispersion-relation}b. The wave modes below $\Omega_e$ are whistler waves. The white solid line in Figure \ref{fig5-dispersion-relation}b represents the dispersion relation of a whistler wave propagating $55^\circ$ with respect to the background magnetic field in a cold plasma. Whistler waves co-streaming with the beam ($k_\parallel > 0$) have a stronger power than the waves counter-streaming with the beam ($k_\parallel < 0$), indicating Landau resonance dominates over cyclotron resonance in the present settings.

\begin{figure}[tphb]
\centering
\includegraphics[width=0.7\textwidth]{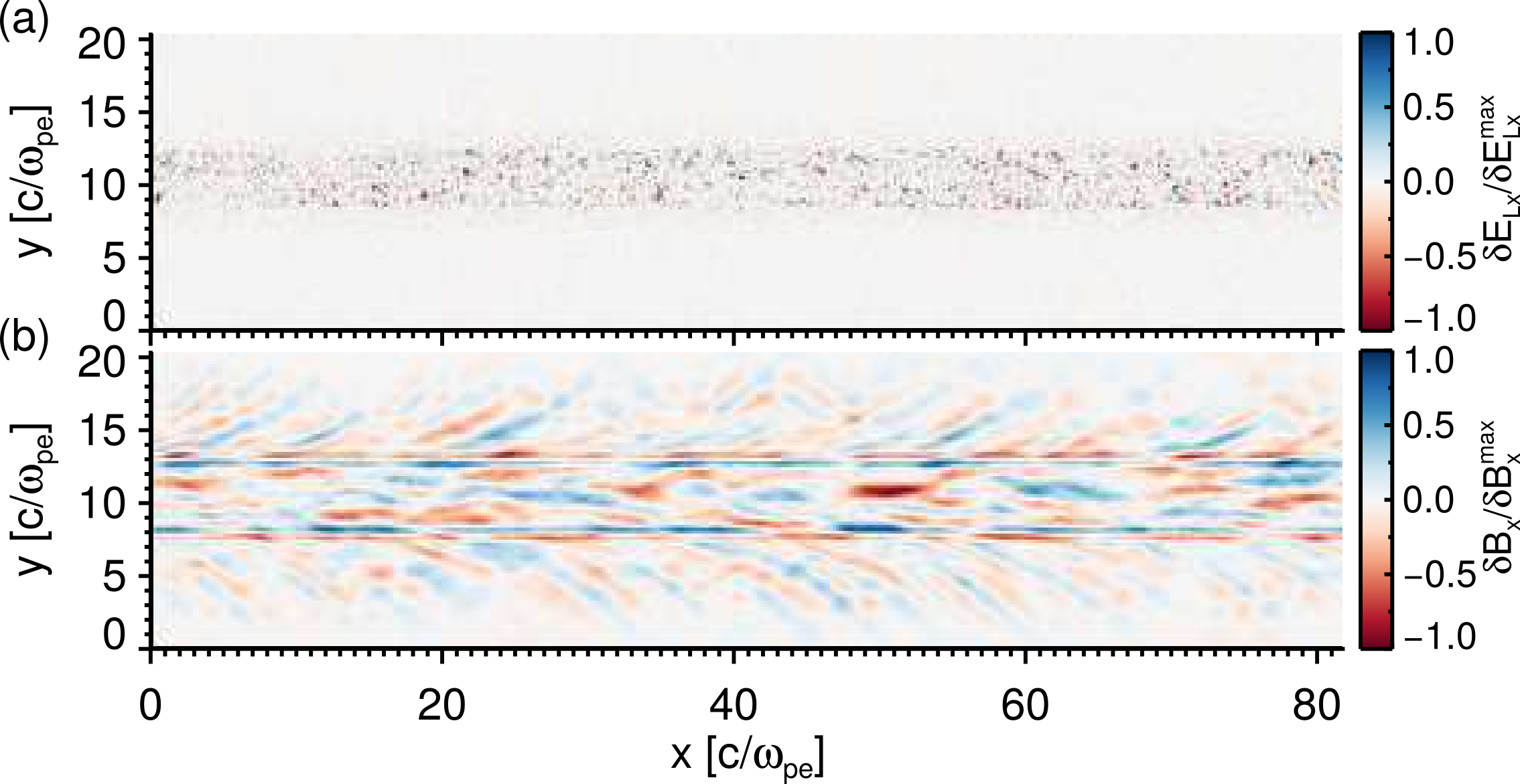}
\caption{(a) The field pattern of the longitudinal electric field along the $x$-direction at $t = 300\, \omega_{pe}^{-1}$. (Multimedia view) (b) The field pattern of the wave magnetic field along the $x$-direction at $t = 300\, \omega_{pe}^{-1}$. (Multimedia view)}
\label{fig5-field-pattern}
\end{figure}

\begin{figure}[tphb]
\centering
\includegraphics[width=0.7\textwidth]{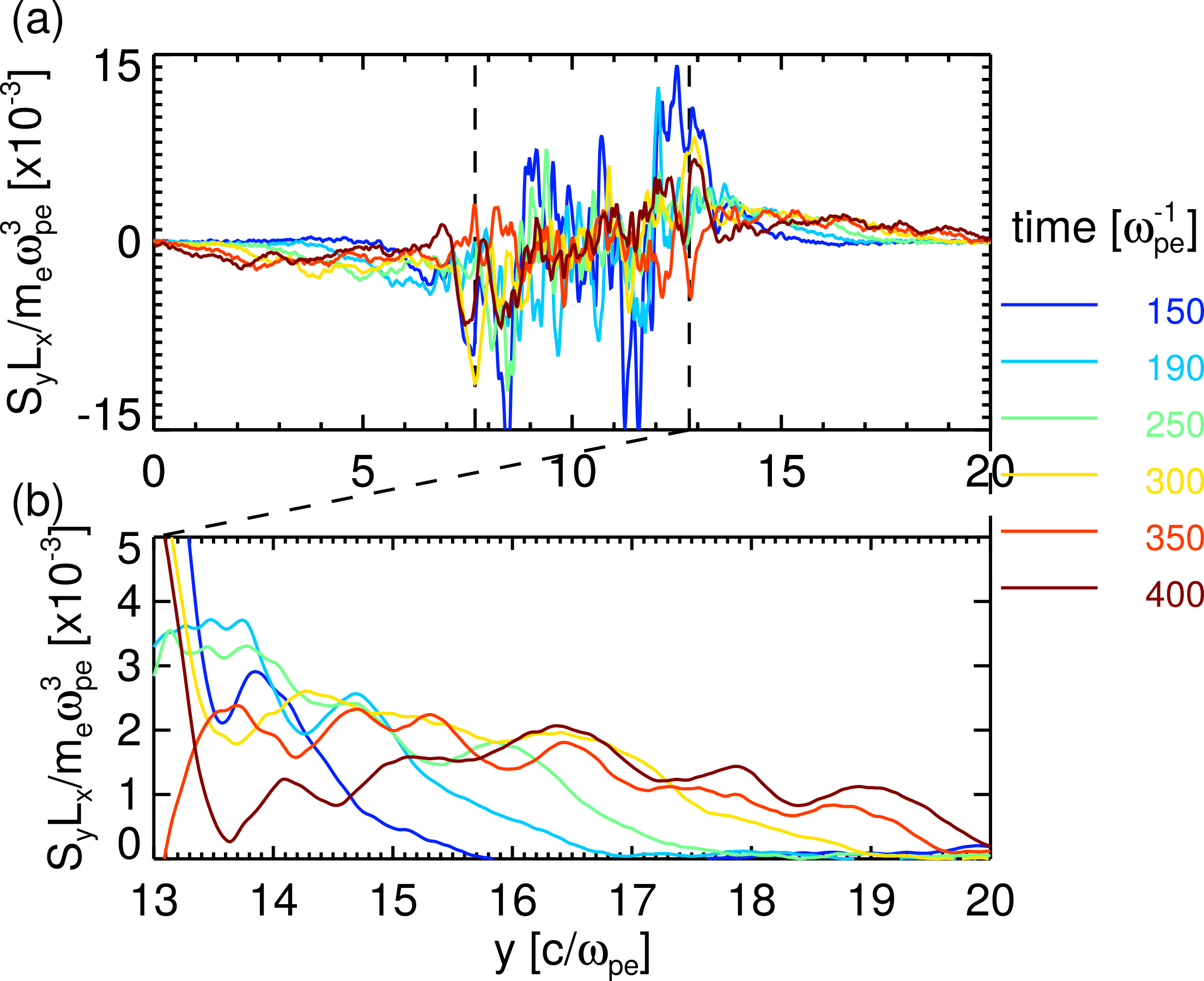}
\caption{(a) The $y$ component of the integrated Poynting flux as a function $y$ position. It is color coded by different time instants corresponding to the legend on the right. The beam region is between the two dashed lines. (b) An expanded display of the integrated Poynting flux for one side out of the beam indicating outflowing energy as a function of time.}
\label{fig5-poynting}
\end{figure}

Figure \ref{fig5-field-pattern} shows the field pattern of electrostatic beam-mode waves and whistler waves in the post-saturation phase. In Figure \ref{fig5-field-pattern}a (Multimedia view), the longitudinal electric field, $\delta\mathbf{E}_L = - \nabla \phi$, along the $x$ direction is displayed at $t = 300\, \omega_{pe}^{-1}$ after electrostatic beam-mode waves saturate. Here $\phi$ represents the electrostatic potential. This field pattern indicate the dominant nature of electrostatic waves at this time, since the electrostatic electric field energy of beam-mode waves is much larger than that of whistler waves. A Fourier analysis of the electrostatic wave field shows that substantial wave energy ranges in the parallel wave number $k_x$ of $10$ - $15\, \omega_{pe}/c$, corresponding to $0.42$ - $0.63\, d_e$ in wavelength. The perpendicular wave number $k_y$ of electrostatic beam-mode waves ranges between $0$ - $4\, \omega_{pe}/c$ at the time of wave saturation, which is much smaller than the parallel wave number $k_x$. It is worthy to note that the excited electrostatic beam-mode waves only exist in the beam region (see the integral multimedia for an animation of the evolution of $\delta E_{Lx}$). In contrast, the excited whistler waves can propagate out of the beam region, as shown by the wave magnetic field $\delta B_x$ in $x$ direction in Figure \ref{fig5-field-pattern}b (Multimedia view). This snapshot is also taken at $t = 300\, \omega_{pe}^{-1}$ after whistler waves saturate. The beam-generated whistler waves have highly oblique wave fronts with $k_x = 1$ - $2\, \omega_{pe}/c$ and $k_y = 1$ - $4\, \omega_{pe}/c$ based on a Fourier analysis of the wave field, corresponding to a wavelength on the order of several electron inertial lengths. It is also noted that there are surface waves at the edge of the beam due to sharp boundaries of the beam density profile (see equation \eqref{eqn:beam-density-profile}). To demonstrate that the energy is flowing out of the beam, the Poynting flux is integrated for all the wave modes along the $x$ direction through the system. Note that the Poynting flux in the Darwin model (see Appendix \ref{append:poynting-flux} for details) differs from that in the electromagnetic model, i.e.,
\begin{eqnarray}\label{eqn:poynting-flux}
\mathbf{S} = \frac{c}{4 \pi} \left[ (\mathbf{E}_L + \mathbf{E}_T)\times \mathbf{B} - \frac{1}{c}\mathbf{E}_T \frac{\partial \phi}{\partial t} \right]
\end{eqnarray}
Here $\mathbf{E}_L$ and $\mathbf{E}_T$ are the longitudinal and transverse components of electric field, respectively, satisfying $\mathbf{\nabla} \times \mathbf{E}_L = 0$ and $\mathbf{\nabla} \cdot \mathbf{E}_T = 0$. The $y$ component of the integrated Poynting flux is shown in Figure \ref{fig5-poynting}. Inside the beam, the Poynting flux can be oriented in both the $+y$ and $-y$ directions, while outside the beam, it is directed only away from the beam indicating that the energy is flowing out of the beam. The region outside of the beam in Figure \ref{fig5-poynting}a is expanded and shown in Figure \ref{fig5-poynting}b. It is seen that the leading edge of the Poynting flux propagates away from the beam as time advances.

%\begin{widetext}
%\begin{center}
%\end{center}
%\end{widetext}

\section{THE EXCITATION OF ELECTROSTATIC BEAM-MODE AND WHISTLER-MODE WAVES AND THE ASSOCIATED EVOLUTION OF THE ELECTRON DISTRIBUTION}
\begin{figure}[tphb]
\centering
\includegraphics[width=0.7\textwidth]{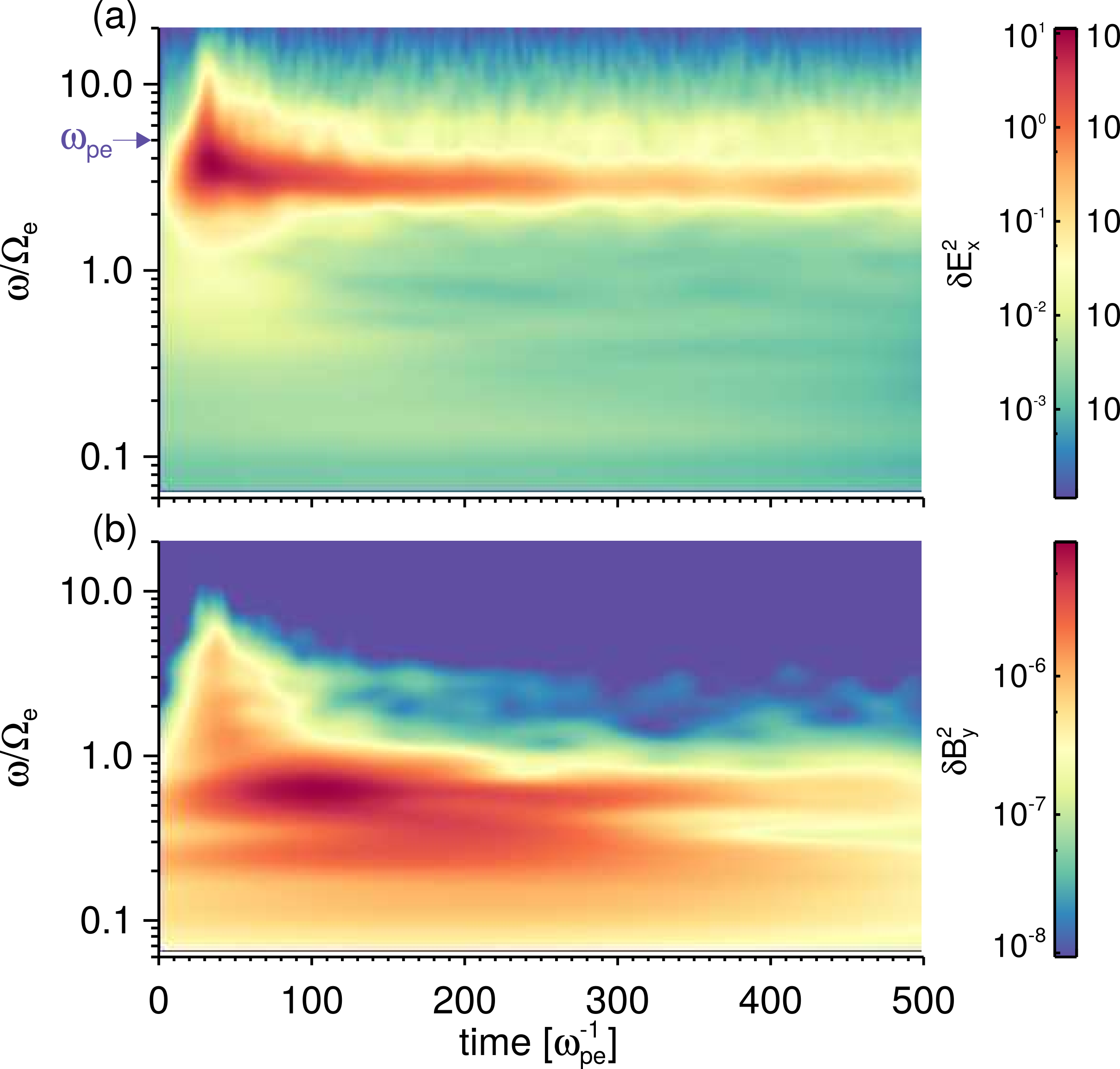}
\caption{(a) The power spectrum of $\delta E_x$ evolving as a function of time. (b) The power spectrum of $\delta B_y$ evolving as a function of time.}
\label{fig5-nominal-wavelet}
\end{figure}

\begin{figure}[tphb]
\centering
\includegraphics[width=0.7\textwidth]{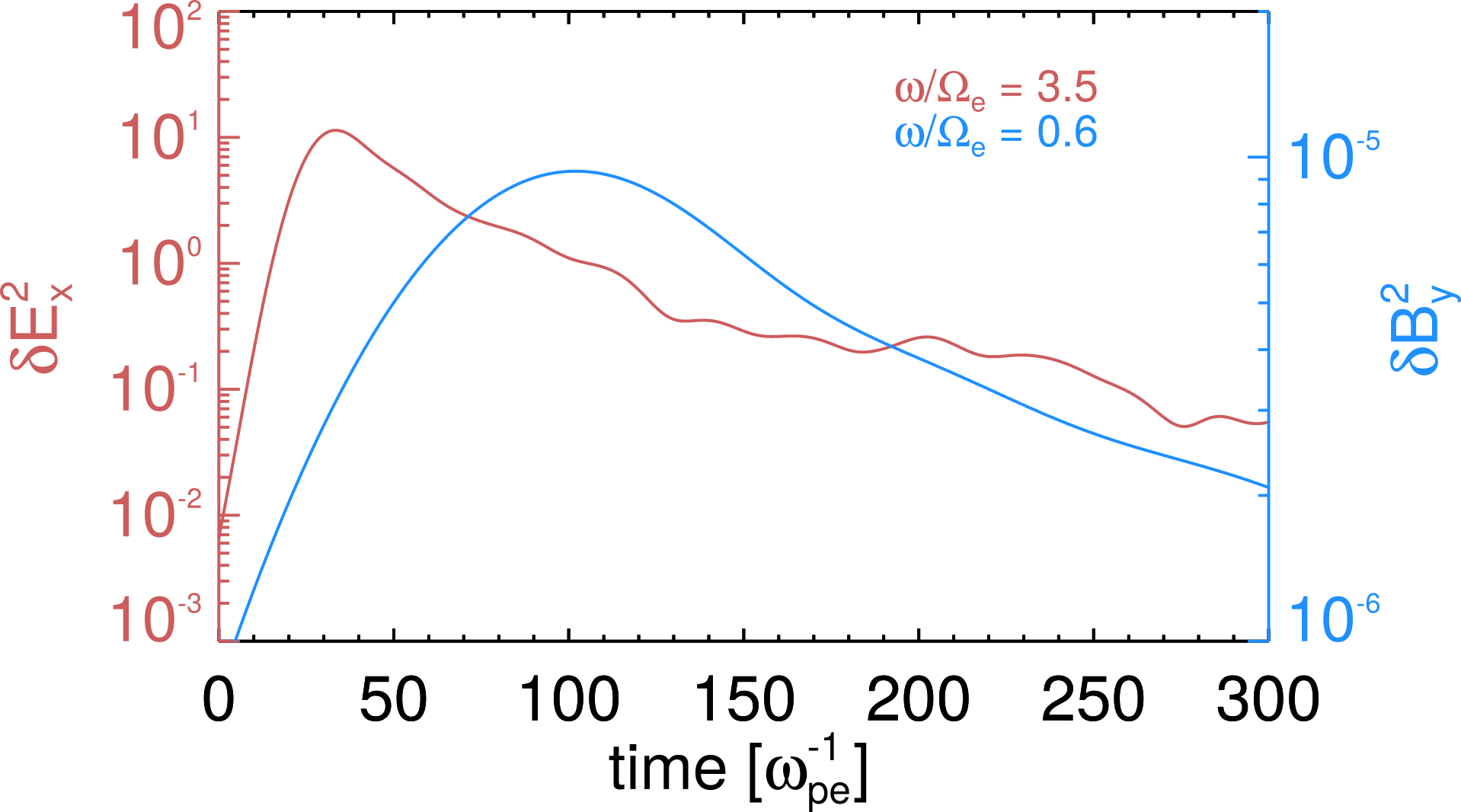}
\caption{The evolution of the power spectral density as a function of time. The power spectral density of $\delta E_x$ at $\omega/\Omega_e = 3.5$ is shown as the red line with the $y$ axis on the left. The power spectral density of $\delta B_y$ at $\omega/\Omega_e = 0.6$ is shown as the blue line with the $y$ axis on the right.}
\label{fig5-nominal-wavelet-cut}
\end{figure}

\begin{figure}[tphb]
\centering
\includegraphics[width=0.7\textwidth]{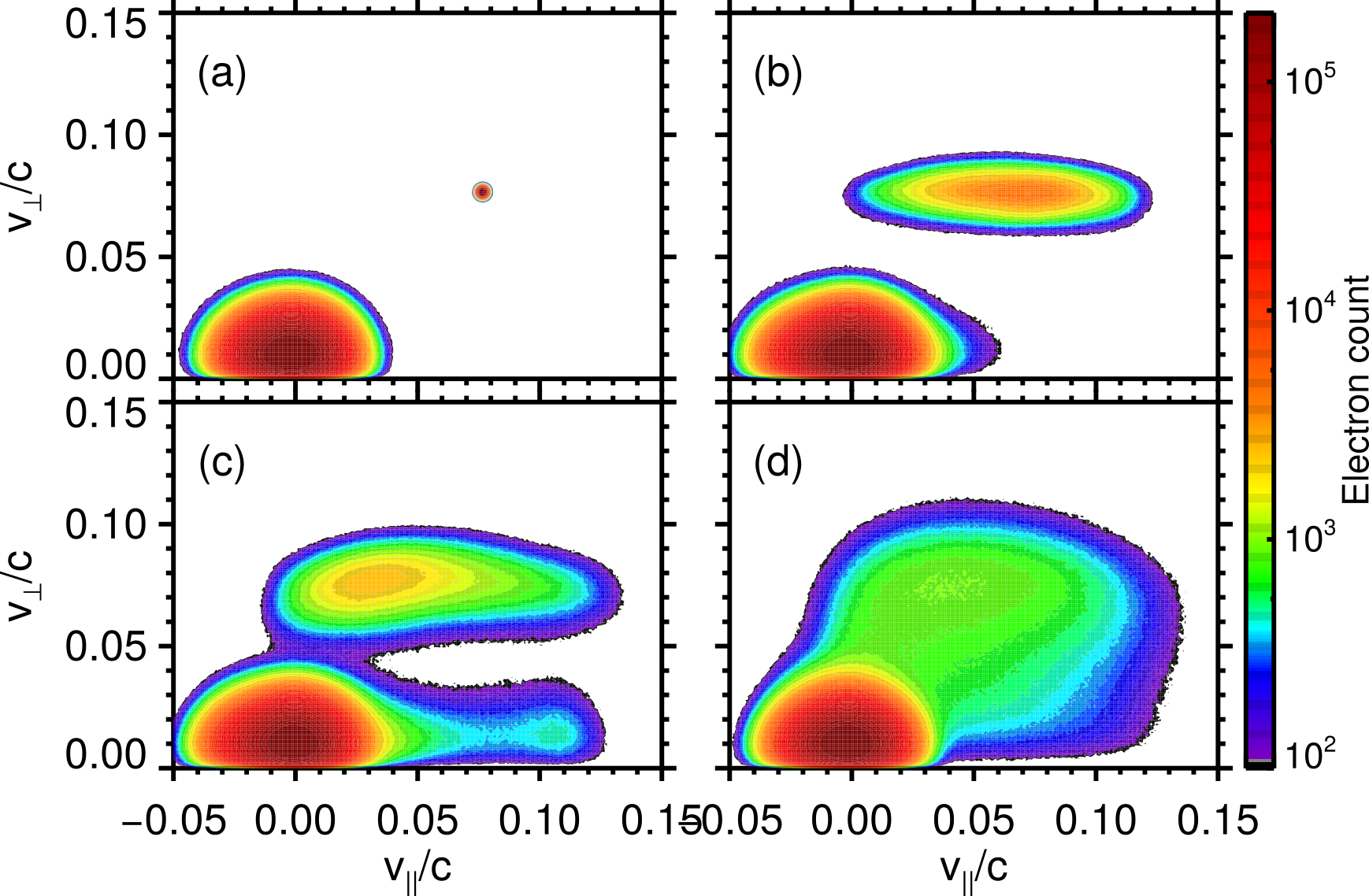}
\caption{The electron distribution in velocity space of $v_\parallel$ - $v_\perp$ at four selected time instants: (a) $t = 0$; (b) $t = 28\, \omega_{pe}^{-1}$; (c) $t = 35\, \omega_{pe}^{-1}$; (d) $t = 100\, \omega_{pe}^{-1}$. See the integral multimedia for an animation of the evolution of the electron distribution. (Multimedia view)}
\label{fig5-electron-distribution}
\end{figure}

We are now in a position to explore the excitation of electrostatic beam-mode and whistler waves and the associated evolution of the electron distribution. The time series data of the electromagnetic fields is sampled at $32$ locations centered in the $x$ direction and equally spaced in the $y$ direction inside the electron beam. A continuous wavelet transform which uses the Morlet wavelet function \citep{Grossmann1984-JMA, Goupillaud1984-Geoeploration}, is applied to the time series data of both the parallel electric field $\delta E_x$ and the $y$ component of the magnetic field $\delta B_y$. The results are shown in Figures \ref{fig5-nominal-wavelet}a and \ref{fig5-nominal-wavelet}b for $\delta E_x$ and $\delta B_y$, respectively. Note that the power spectrum is averaged over $32$ sampling locations to minimize its variance. Electrostatic beam-mode waves at $\omega/\Omega_e = 3$ - $5$ dominate over other wave modes in the power spectrum of $\delta E_x$ as shown in Figure \ref{fig5-nominal-wavelet}a. They saturate in approximately five plasma oscillations (around $t = 30\, \omega_{pe}^{-1}$) and gradually damp out. Whistler waves show up prominently below the electron cyclotron frequency in the power spectrum of $\delta B_y$. Around $t = 100\, \omega_{pe}^{-1}$ ($\sim 3$ cyclotron periods), whistler waves saturate with a primary peak at $\omega/\Omega_e = 0.6$ and a secondary peak at $\omega/\Omega_e = 0.25$. After saturation, the magnitude of these oblique whistler waves further decreases through Landau damping. To contrast the very different growth rates between electrostatic beam-mode waves compared to whistler waves, two line cuts are taken from the wavelet spectral peaks, one at $\omega/\Omega_e = 3.5$ for electrostatic beam-mode waves and the other at $\omega/\Omega_e = 0.6$ for whistler waves. The results are shown in a linear-log plot in Figure \ref{fig5-nominal-wavelet-cut}. The magnitude of the linear growth rate corresponds to $1/2$ of the slope in the linear part of the wave energy evolution. This linear growth rate is calculated to be $0.15\, \omega_{pe}$ for electrostatic beam-mode waves at $\omega/\Omega_e = 3.5$, and $0.015\, \omega_{pe}$ ($= 0.075\, \Omega_e$) for whistler waves at $\omega/\Omega_e = 0.6$. This calculation characterizes the rapidly growing electrostatic beam-mode waves and relatively slow-growing whistler waves. Note that before the electrostatic beam-mode wave saturates, whistler waves can also extract free energy from the inverted slope region (i.e., $\partial f_b / \partial v_\parallel > 0$) of the beam through Landau resonance, although the rate of such energy transfer is slower than that for the electrostatic beam-mode wave as shown in Figure \ref{fig5-nominal-wavelet-cut}. After the electrostatic beam-mode wave saturates, whistler waves can only be excited through cyclotron resonance since the free energy from $\partial f_b / \partial v_\parallel > 0$ has been exhausted by the electrostatic instability. Correspondingly, the electron distribution responds to the electrostatic and whistler instabilities on two different time scales. Figure \ref{fig5-electron-distribution} (Multimedia view) shows the electron distribution in velocity space, $v_\parallel$ - $v_\perp$ at four representative times. Note that the electrons are counted over the entire computation domain. To begin, the distribution is initialized with a population of core electrons and a separate population of beam ring electrons (Figure \ref{fig5-electron-distribution}a). Shortly before the electrostatic beam-mode wave saturation at $t = 28\, \omega_{pe}^{-1}$, the beam electrons are trapped and relaxed by the electrostatic beam-mode waves in the parallel direction (Figure \ref{fig5-electron-distribution}b). As the magnitude of the electrostatic beam-mode wave grows, the width of its resonant island broadens in $v_\parallel$ due to $\Delta v_\parallel \propto \sqrt{\delta E}$, where $\Delta v_\parallel$ is the width of the resonant island and $\delta E$ is the electrostatic beam-mode wave amplitude. This large amplitude electrostatic wave becomes resonant with, and traps the tail of the core electrons and subsequently gets the tail of the core electrons accelerated to the beam energy level, as shown in Figure \ref{fig5-electron-distribution}c at $t = 35\, \omega_{pe}^{-1}$. At a later time, the relaxed beam electrons are scattered along resonant diffusion surfaces to lower pitch angles and lose energy, through which whistler waves further gain energy and grow in magnitude. This is shown in Figure \ref{fig5-electron-distribution}d taken at $t = 100\, \omega_{pe}^{-1}$ when the whistler waves saturate.

\section{The suppression of beam whistler instabilities by electrostatic beam-mode wave}
\begin{figure}[tphb]
\centering
\includegraphics[width=0.7\textwidth]{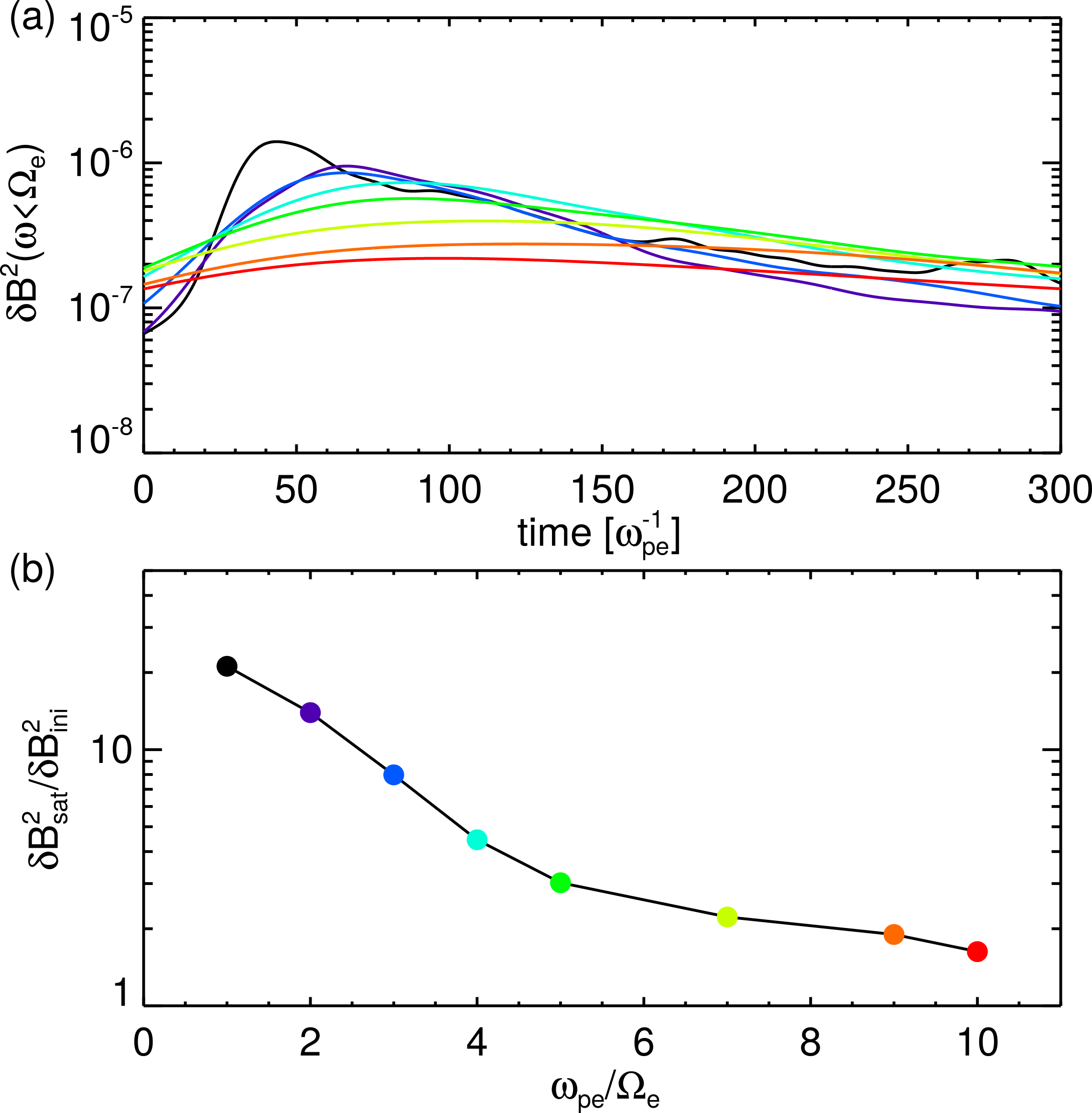}
\caption{(a) the evolution of magnetic field energy of whistler waves as a function of time. Starting from the black-blue line and going to the orange-red line, the corresponding values of $\omega_{pe}/\Omega_e$ are $1,2,3,4,5,7,9,10$. (b) Corresponding to each run in (a), the ratio of saturated energy to initial energy is shown as a function of $\omega_{pe}/\Omega_e$. Each colored spot corresponds to the line of the same color in panel (a).}
\label{fig5-fpefce-dependence}
\end{figure}

The growth of whistler-mode waves through Landau resonance is limited by the growth of electrostatic beam-mode waves. The fast growing electrostatic waves saturate rapidly in a few plasma oscillations and deplete the beam free energy in the parallel direction through Landau resonance. Whistler waves saturate soon after the saturation of electrostatic beam-mode waves since there is little free energy left for the Landau resonant excitation of whistler waves. Such a competition between electrostatic and whistler instabilities depends on $\omega_{pe}/\Omega_e$, which characterizes the ratio between the linear growth rate of electrostatic instabilities and that of whistler instabilities. To test this idea and minimize the effect of cyclotron resonance, a field-aligned electron beam is used here while the rest of the setup is kept the same. Figure \ref{fig5-fpefce-dependence}a shows the magnetic field energy of whistler waves with respect to time for a set of $\omega_{pe}/\Omega_e$ values. Each of the color-coded lines corresponds to the colored spot in Figure \ref{fig5-fpefce-dependence}b, in which the ratio of the saturated magnetic field energy to initial magnetic field energy is shown as a function of $\omega_{pe}/\Omega_e$. Under the special scenario of $\omega_{pe}/\Omega_e = 1$, whistler waves and electrostatic beam-mode waves saturate over the same time scale and whistler waves saturate at a substantially larger amplitude compared to other cases. As $\omega_{pe}/\Omega_e$ increases, the saturated whistler wave energy decreases and eventually is immersed in the noise level beyond $\omega_{pe}/\Omega_e = 7$. Linear theory predicts that Landau resonance between whistler waves and the electron beam does not occur beyond a critical value of $(\omega_{pe}/\Omega_e)_{critical} = 6.5$ for a cold beam in our parameter regime (see Appendix A for details). This inhibits the energy transfer between the beam electrons and whistler waves and results in a low signal to noise ratio in the high $\omega_{pe}/\Omega_e$ regime. Below the critical value of $\omega_{pe}/\Omega_e = 6.5$, electrostatic instabilities limit the saturation energy level of whistler instabilities by extracting the free energy of the beam at a faster rate than the whistler instabilities as long as $\omega_{pe}/\Omega_e > 1$. It is also noted that there is a weak trend of decreasing signal to noise ratio beyond the critical value of $\omega_{pe}/\Omega_e = 6.5$. This may result from the fact that the theory prediction is for a cold beam while the distribution function is relaxed from the cold beam ring in the kinetic simulations and therefore it leads to a weak energy transfer between beam electrons and whistler waves even beyond the predicted critical value.

\section{Summary and discussion}
Using a self-consistent Darwin particle-in-cell method, we study the excitation of electrostatic beam-mode and whistler waves in a beam-plasma system. The electrostatic beam-mode waves grow in magnitude rapidly and saturate in a few plasma oscillations, while the electron beam is slowed down and relaxed in the parallel direction. As the amplitude of electrostatic beam-mode waves approaches saturation, resonance with the tail of the background core electrons occurs and accelerates them parallel to the background magnetic field. Whistler waves grow in magnitude and saturate over the time scale of a few cyclotron periods. They are excited through Landau resonance and cyclotron resonance. In terms of wave propagation, electrostatic beam-mode waves are localized to the beam region, whereas whistler waves can leak out of the beam and transport energy away from the beam. Finally, the competition between electrostatic and whistler instabilities are tested for a field-aligned beam. Due to a faster depletion of the beam free energy by electrostatic beam-mode waves with increasing $\omega_{pe}/\Omega_e$, the saturation amplitude of whistler waves decreases. Beyond a critical $\omega_{pe}/\Omega_e$, Landau resonance does not occur for whistler waves and the saturation amplitude of whistler waves is immersed in the noise.\\

There are still a number of differences between the kinetic simulation results and observations made with laboratory experiments. First, the PIC simulation is a relaxation of an initial beam whereas in the experiment, the beam electrons are continuously injected into the cold plasma. Second, the PIC simulation has the beam uniformly distributed along the parallel direction whereas in the experiment the beam source is fixed at a specific location along the parallel direction. In the experiment, it takes about $40$ cyclotron periods or $200$ plasma oscillations for the beam electrons to travel from the source location to the end of the experimental volume. In other words, the transit time of the beam electrons is $\Delta t \approx 1257\, \omega_{pe}^{-1}$. In the PIC simulation with the ratio of beam density to total plasma density as $n_b/n_t = 0.125$, electrostatic beam-mode and whistler waves, respectively, saturate at $t = 30\, \omega_{pe}^{-1}$ and $t = 100\, \omega_{pe}^{-1}$. However, the saturation time of waves with the simulation value $n_b/n_t = 0.125$ should be properly scaled to the experimental value of $n_b/n_t = 0.001 \sim 0.005$. As a rough estimate, suppose that the saturation time of waves is inversely proportional to the linear growth rate of waves, i.e., $t_{sat} \propto 1/\gamma$, and that the linear growth rate of waves scales with the beam density \citep{Oneil1971-PoF, Bell1964-PR} as $\gamma \propto (n_b/n_t)^{\frac{1}{3}}$ and hence $t_{sat} \propto (n_b/n_t)^{-\frac{1}{3}}$. Electrostatic beam-mode and whistler waves, respectively, are estimated to saturate at $t = 90 \sim 150\, \omega_{pe}^{-1}$ and $t = 300 \sim 500\, \omega_{pe}^{-1}$ in the experiment. Therefore it is expected that the electron distribution would be fully relaxed as in Figure \ref{fig5-electron-distribution}d at the end of the experimental volume with a transit time of $\Delta t \approx 1257\, \omega_{pe}^{-1}$ in the experiment. The more realistic situation of the injection experiment driven by a beam source will be implemented in the PIC simulation in a future study.

\appendix
\section{A critical value of $\omega_{pe}/\Omega_e$ for Landau resonance between whistler waves and beam electrons}
It can be shown that \citep{Starodubtsev1999-PoP} Landau resonance between whistler waves and beam electrons only occurs below some critical $\omega_{pe}/\Omega_e$. From the refractive index surface of whistler waves, there exists a minimum $k_z$ for a given frequency 
\begin{eqnarray}
k_z^{\min} = \begin{cases}
\dfrac{\omega_{pe}}{c}\dfrac{2\omega}{\Omega_e} & \omega < \dfrac{\Omega_e}{2} \\
\dfrac{\omega_{pe}}{c} \sqrt{\dfrac{\omega}{\Omega_e-\omega}} & \omega > \dfrac{\Omega_e}{2}
\end{cases}
\end{eqnarray}
Note that $k_z$ takes the minimum value at the Gendrin angle for $\omega < \Omega_e/2$, while for $\omega > \Omega_e/2$, $k_z$ takes the minimum value in the parallel direction. In order to have Landau resonance between beam electrons and whistler waves, the resonant wave number must exceed $k_z^{\min}$. That is 
\begin{eqnarray}
\frac{\omega}{v_{z}} > k_z^{\min}
\end{eqnarray}
There exists a critical value of $\omega_{pe}/\Omega_e$, above which Landau resonance does not occur. This critical value is
\begin{eqnarray}
\left(\frac{\omega_{pe}}{\Omega_e}\right)_{\mbox{critical}} = \begin{cases}
\frac{c}{2v_{z}} & \omega < \frac{\Omega_e}{2} \\
\sqrt{\frac{\omega}{\Omega_e} \left(1-\frac{\omega}{\Omega_e}\right)}\frac{c}{v_{z}} & \omega > \frac{\Omega_e}{2}
\end{cases}
\end{eqnarray}
For typical parameters in the simulation, i.e., $v_z/c=0.0766$ and $\omega/\Omega_e=0.5$, the critical value of $\omega_{pe}/\Omega_e$ is $6.5$.

\section{The equation of energy flux in the Darwin model}\label{append:poynting-flux}
The energy flux equation in the Darwin model is
\begin{eqnarray}\label{eqn:energy-flux}
\mathbf{\nabla} \cdot \mathbf{S} + \frac{\partial}{\partial t} \left[ \frac{\mathbf{E}_L \cdot \mathbf{E}_L}{8 \pi} + \frac{\mathbf{B} \cdot \mathbf{B}}{8 \pi} \right] = - \mathbf{J} \cdot (\mathbf{E}_L + \mathbf{E}_T)
\end{eqnarray}
where the Poynting flux takes the form of equation \eqref{eqn:poynting-flux}. $\mathbf{J}$ is the current density. Note that the energy of transverse electric field does not enter the field energy. The energy flux equation can be verified immediately by taking divergence of the Poynting flux in equation \eqref{eqn:poynting-flux} and making use of the following set of equations
\begin{eqnarray}
\mathbf{\nabla} \times \mathbf{B} &=& \frac{4 \pi}{c} \mathbf{J} + \frac{1}{c} \frac{\partial \mathbf{E}_L}{\partial t} \label{eqn:curlB} \\
\mathbf{\nabla} \times \mathbf{E}_T &=& -\frac{1}{c} \frac{\partial \mathbf{B}}{\partial t} \\
\mathbf{\nabla} \cdot \mathbf{B} &=& 0 \\
\mathbf{\nabla} \cdot \mathbf{E}_T &=& 0 \\
\mathbf{\nabla} \times \mathbf{E}_L &=& 0 
\end{eqnarray}
The transverse component of the displacement current is neglected in equation \eqref{eqn:curlB} due to the Darwin approximation.

% If in two-column mode, this environment will change to single-column format so that long equations can be displayed. 
%\begin{widetext}
%$$\mbox{put long equation here}$$
%\end{widetext}

% Figures should be put into the text as floats. 
% Use the graphics or graphicx packages (distributed with LaTeX2e).
% Put the label that you will use with \ref{} command in the braces of the \label{} command.
%
% \begin{figure}
% \end{figure}

% \begin{table}
% \end{tabular}
% \end{table}

\begin{acknowledgments}
We thank G. J. Morales for helpful discussions. We would also like to acknowledge high-performance computing support from Yellowstone (ark:/85065/d7wd3xhc) provided by NCAR's Computational and Information Systems Laboratory, sponsored by the National Science Foundation. The research was funded by the Department of Energy and the National Science Foundation by grant DE-SC0010578, which was awarded to UCLA through the NSF/DOE Plasma Partnership program. The research was also funded by NASA grant NNX16AG21G.
\end{acknowledgments}

%\bibliography{your-bib-file}
\bibliography{LangWhist_pic}

\begin{thebibliography}{43}%
\makeatletter
\providecommand \@ifxundefined [1]{%
 \@ifx{#1\undefined}
}%
\providecommand \@ifnum [1]{%
 \ifnum #1\expandafter \@firstoftwo
 \else \expandafter \@secondoftwo
 \fi
}%
\providecommand \@ifx [1]{%
 \ifx #1\expandafter \@firstoftwo
 \else \expandafter \@secondoftwo
 \fi
}%
\providecommand \natexlab [1]{#1}%
\providecommand \enquote  [1]{``#1''}%
\providecommand \bibnamefont  [1]{#1}%
\providecommand \bibfnamefont [1]{#1}%
\providecommand \citenamefont [1]{#1}%
\providecommand \href@noop [0]{\@secondoftwo}%
\providecommand \href [0]{\begingroup \@sanitize@url \@href}%
\providecommand \@href[1]{\@@startlink{#1}\@@href}%
\providecommand \@@href[1]{\endgroup#1\@@endlink}%
\providecommand \@sanitize@url [0]{\catcode `\\12\catcode `\$12\catcode
  `\&12\catcode `\#12\catcode `\^12\catcode `\_12\catcode `\%12\relax}%
\providecommand \@@startlink[1]{}%
\providecommand \@@endlink[0]{}%
\providecommand \url  [0]{\begingroup\@sanitize@url \@url }%
\providecommand \@url [1]{\endgroup\@href {#1}{\urlprefix }}%
\providecommand \urlprefix  [0]{URL }%
\providecommand \Eprint [0]{\href }%
\providecommand \doibase [0]{http://dx.doi.org/}%
\providecommand \selectlanguage [0]{\@gobble}%
\providecommand \bibinfo  [0]{\@secondoftwo}%
\providecommand \bibfield  [0]{\@secondoftwo}%
\providecommand \translation [1]{[#1]}%
\providecommand \BibitemOpen [0]{}%
\providecommand \bibitemStop [0]{}%
\providecommand \bibitemNoStop [0]{.\EOS\space}%
\providecommand \EOS [0]{\spacefactor3000\relax}%
\providecommand \BibitemShut  [1]{\csname bibitem#1\endcsname}%
\let\auto@bib@innerbib\@empty
%</preamble>
\bibitem [{\citenamefont {Tokar}, \citenamefont {Gurnett},\ and\ \citenamefont
  {Feldman}(1984)}]{Tokar1984-JGR}%
  \BibitemOpen
  \bibfield  {author} {\bibinfo {author} {\bibfnamefont {R.~L.}\ \bibnamefont
  {Tokar}}, \bibinfo {author} {\bibfnamefont {D.~A.}\ \bibnamefont {Gurnett}},
  \ and\ \bibinfo {author} {\bibfnamefont {W.~C.}\ \bibnamefont {Feldman}},\
  }\bibfield  {title} {\enquote {\bibinfo {title} {Whistler mode turbulence
  generated by electron beams in earth's bow shock},}\ }\href {\doibase
  10.1029/JA089iA01p00105} {\bibfield  {journal} {\bibinfo  {journal} {Journal
  of Geophysical Research: Space Physics}\ }\textbf {\bibinfo {volume} {89}},\
  \bibinfo {pages} {105--114} (\bibinfo {year} {1984})}\BibitemShut {NoStop}%
\bibitem [{\citenamefont {Marsch}(1985)}]{Marsch1985-JGR}%
  \BibitemOpen
  \bibfield  {author} {\bibinfo {author} {\bibfnamefont {E.}~\bibnamefont
  {Marsch}},\ }\bibfield  {title} {\enquote {\bibinfo {title} {Beam-driven
  electron acoustic waves upstream of the earth's bow shock},}\ }\href
  {\doibase 10.1029/JA090iA07p06327} {\bibfield  {journal} {\bibinfo  {journal}
  {Journal of Geophysical Research: Space Physics}\ }\textbf {\bibinfo {volume}
  {90}},\ \bibinfo {pages} {6327--6336} (\bibinfo {year} {1985})}\BibitemShut
  {NoStop}%
\bibitem [{\citenamefont {Bale}\ \emph {et~al.}(1999)\citenamefont {Bale},
  \citenamefont {Reiner}, \citenamefont {Bougeret}, \citenamefont {Kaiser},
  \citenamefont {Krucker}, \citenamefont {Larson},\ and\ \citenamefont
  {Lin}}]{Bale1999-GRL}%
  \BibitemOpen
  \bibfield  {author} {\bibinfo {author} {\bibfnamefont {S.~D.}\ \bibnamefont
  {Bale}}, \bibinfo {author} {\bibfnamefont {M.~J.}\ \bibnamefont {Reiner}},
  \bibinfo {author} {\bibfnamefont {J.~L.}\ \bibnamefont {Bougeret}}, \bibinfo
  {author} {\bibfnamefont {M.~L.}\ \bibnamefont {Kaiser}}, \bibinfo {author}
  {\bibfnamefont {S.}~\bibnamefont {Krucker}}, \bibinfo {author} {\bibfnamefont
  {D.~E.}\ \bibnamefont {Larson}}, \ and\ \bibinfo {author} {\bibfnamefont
  {R.~P.}\ \bibnamefont {Lin}},\ }\bibfield  {title} {\enquote {\bibinfo
  {title} {The source region of an interplanetary type {II} radio burst},}\
  }\href {\doibase 10.1029/1999GL900293} {\bibfield  {journal} {\bibinfo
  {journal} {Geophysical Research Letters}\ }\textbf {\bibinfo {volume} {26}},\
  \bibinfo {pages} {1573--1576} (\bibinfo {year} {1999})}\BibitemShut {NoStop}%
\bibitem [{\citenamefont {Carlson}\ \emph {et~al.}(1998)\citenamefont
  {Carlson}, \citenamefont {McFadden}, \citenamefont {Ergun}, \citenamefont
  {Temerin}, \citenamefont {Peria}, \citenamefont {Mozer}, \citenamefont
  {Klumpar}, \citenamefont {Shelley}, \citenamefont {Peterson}, \citenamefont
  {Moebius}, \citenamefont {Elphic}, \citenamefont {Strangeway}, \citenamefont
  {Cattell},\ and\ \citenamefont {Pfaff}}]{Carlson1998-GRL}%
  \BibitemOpen
  \bibfield  {author} {\bibinfo {author} {\bibfnamefont {C.~W.}\ \bibnamefont
  {Carlson}}, \bibinfo {author} {\bibfnamefont {J.~P.}\ \bibnamefont
  {McFadden}}, \bibinfo {author} {\bibfnamefont {R.~E.}\ \bibnamefont {Ergun}},
  \bibinfo {author} {\bibfnamefont {M.}~\bibnamefont {Temerin}}, \bibinfo
  {author} {\bibfnamefont {W.}~\bibnamefont {Peria}}, \bibinfo {author}
  {\bibfnamefont {F.~S.}\ \bibnamefont {Mozer}}, \bibinfo {author}
  {\bibfnamefont {D.~M.}\ \bibnamefont {Klumpar}}, \bibinfo {author}
  {\bibfnamefont {E.~G.}\ \bibnamefont {Shelley}}, \bibinfo {author}
  {\bibfnamefont {W.~K.}\ \bibnamefont {Peterson}}, \bibinfo {author}
  {\bibfnamefont {E.}~\bibnamefont {Moebius}}, \bibinfo {author} {\bibfnamefont
  {R.}~\bibnamefont {Elphic}}, \bibinfo {author} {\bibfnamefont
  {R.}~\bibnamefont {Strangeway}}, \bibinfo {author} {\bibfnamefont
  {C.}~\bibnamefont {Cattell}}, \ and\ \bibinfo {author} {\bibfnamefont
  {R.}~\bibnamefont {Pfaff}},\ }\bibfield  {title} {\enquote {\bibinfo {title}
  {Fast observations in the downward auroral current region: Energetic upgoing
  electron beams, parallel potential drops, and ion heating},}\ }\href
  {\doibase 10.1029/98GL00851} {\bibfield  {journal} {\bibinfo  {journal}
  {Geophysical Research Letters}\ }\textbf {\bibinfo {volume} {25}},\ \bibinfo
  {pages} {2017--2020} (\bibinfo {year} {1998})}\BibitemShut {NoStop}%
\bibitem [{\citenamefont {Maggs}(1978)}]{Maggs1978-JGR}%
  \BibitemOpen
  \bibfield  {author} {\bibinfo {author} {\bibfnamefont {J.~E.}\ \bibnamefont
  {Maggs}},\ }\bibfield  {title} {\enquote {\bibinfo {title} {Electrostatic
  noise generated by the auroral electron beam},}\ }\href {\doibase
  10.1029/JA083iA07p03173} {\bibfield  {journal} {\bibinfo  {journal} {Journal
  of Geophysical Research: Space Physics}\ }\textbf {\bibinfo {volume} {83}},\
  \bibinfo {pages} {3173--3188} (\bibinfo {year} {1978})}\BibitemShut {NoStop}%
\bibitem [{\citenamefont {Petrosian}(1973)}]{petrosian1973impulsive}%
  \BibitemOpen
  \bibfield  {author} {\bibinfo {author} {\bibfnamefont {V.}~\bibnamefont
  {Petrosian}},\ }\bibfield  {title} {\enquote {\bibinfo {title} {Impulsive
  solar x-ray bursts: bremsstrahlung radiation from a beam of electrons in the
  solar chromosphere and the total energy of solar flares},}\ }\href@noop {}
  {\bibfield  {journal} {\bibinfo  {journal} {The Astrophysical Journal}\
  }\textbf {\bibinfo {volume} {186}},\ \bibinfo {pages} {291--304} (\bibinfo
  {year} {1973})}\BibitemShut {NoStop}%
\bibitem [{\citenamefont {Drake}\ \emph {et~al.}(2003)\citenamefont {Drake},
  \citenamefont {Swisdak}, \citenamefont {Cattell}, \citenamefont {Shay},
  \citenamefont {Rogers},\ and\ \citenamefont {Zeiler}}]{drake2003formation}%
  \BibitemOpen
  \bibfield  {author} {\bibinfo {author} {\bibfnamefont {J.}~\bibnamefont
  {Drake}}, \bibinfo {author} {\bibfnamefont {M.}~\bibnamefont {Swisdak}},
  \bibinfo {author} {\bibfnamefont {C.}~\bibnamefont {Cattell}}, \bibinfo
  {author} {\bibfnamefont {M.}~\bibnamefont {Shay}}, \bibinfo {author}
  {\bibfnamefont {B.}~\bibnamefont {Rogers}}, \ and\ \bibinfo {author}
  {\bibfnamefont {A.}~\bibnamefont {Zeiler}},\ }\bibfield  {title} {\enquote
  {\bibinfo {title} {Formation of electron holes and particle energization
  during magnetic reconnection},}\ }\href@noop {} {\bibfield  {journal}
  {\bibinfo  {journal} {Science}\ }\textbf {\bibinfo {volume} {299}},\ \bibinfo
  {pages} {873--877} (\bibinfo {year} {2003})}\BibitemShut {NoStop}%
\bibitem [{\citenamefont {Pritchett}\ and\ \citenamefont
  {Coroniti}(2004)}]{Pritchett2004-JGR}%
  \BibitemOpen
  \bibfield  {author} {\bibinfo {author} {\bibfnamefont {P.~L.}\ \bibnamefont
  {Pritchett}}\ and\ \bibinfo {author} {\bibfnamefont {F.~V.}\ \bibnamefont
  {Coroniti}},\ }\bibfield  {title} {\enquote {\bibinfo {title}
  {Three-dimensional collisionless magnetic reconnection in the presence of a
  guide field},}\ }\href {\doibase 10.1029/2003JA009999} {\bibfield  {journal}
  {\bibinfo  {journal} {Journal of Geophysical Research: Space Physics}\
  }\textbf {\bibinfo {volume} {109}},\ \bibinfo {pages} {{A}01220} (\bibinfo
  {year} {2004})}\BibitemShut {NoStop}%
\bibitem [{\citenamefont {Li}\ \emph {et~al.}(2016)\citenamefont {Li},
  \citenamefont {Mourenas}, \citenamefont {Artemyev}, \citenamefont {Bortnik},
  \citenamefont {Thorne}, \citenamefont {Kletzing}, \citenamefont {Kurth},
  \citenamefont {Hospodarsky}, \citenamefont {Reeves}, \citenamefont
  {Funsten},\ and\ \citenamefont {Spence}}]{Li2016-GRL}%
  \BibitemOpen
  \bibfield  {author} {\bibinfo {author} {\bibfnamefont {W.}~\bibnamefont
  {Li}}, \bibinfo {author} {\bibfnamefont {D.}~\bibnamefont {Mourenas}},
  \bibinfo {author} {\bibfnamefont {A.~V.}\ \bibnamefont {Artemyev}}, \bibinfo
  {author} {\bibfnamefont {J.}~\bibnamefont {Bortnik}}, \bibinfo {author}
  {\bibfnamefont {R.~M.}\ \bibnamefont {Thorne}}, \bibinfo {author}
  {\bibfnamefont {C.~A.}\ \bibnamefont {Kletzing}}, \bibinfo {author}
  {\bibfnamefont {W.~S.}\ \bibnamefont {Kurth}}, \bibinfo {author}
  {\bibfnamefont {G.~B.}\ \bibnamefont {Hospodarsky}}, \bibinfo {author}
  {\bibfnamefont {G.~D.}\ \bibnamefont {Reeves}}, \bibinfo {author}
  {\bibfnamefont {H.~O.}\ \bibnamefont {Funsten}}, \ and\ \bibinfo {author}
  {\bibfnamefont {H.~E.}\ \bibnamefont {Spence}},\ }\bibfield  {title}
  {\enquote {\bibinfo {title} {Unraveling the excitation mechanisms of highly
  oblique lower band chorus waves},}\ }\href {\doibase 10.1002/2016GL070386}
  {\bibfield  {journal} {\bibinfo  {journal} {Geophysical Research Letters}\
  }\textbf {\bibinfo {volume} {43}},\ \bibinfo {pages} {8867--8875} (\bibinfo
  {year} {2016})},\ \bibinfo {note} {2016GL070386}\BibitemShut {NoStop}%
\bibitem [{\citenamefont {O'Neil}, \citenamefont {Winfrey},\ and\ \citenamefont
  {Malmberg}(1971)}]{Oneil1971-PoF}%
  \BibitemOpen
  \bibfield  {author} {\bibinfo {author} {\bibfnamefont {T.~M.}\ \bibnamefont
  {O'Neil}}, \bibinfo {author} {\bibfnamefont {J.~H.}\ \bibnamefont {Winfrey}},
  \ and\ \bibinfo {author} {\bibfnamefont {J.~H.}\ \bibnamefont {Malmberg}},\
  }\bibfield  {title} {\enquote {\bibinfo {title} {Nonlinear interaction of a
  small cold beam and a plasma},}\ }\href {\doibase 10.1063/1.1693587}
  {\bibfield  {journal} {\bibinfo  {journal} {The Physics of Fluids}\ }\textbf
  {\bibinfo {volume} {14}},\ \bibinfo {pages} {1204--1212} (\bibinfo {year}
  {1971})},\ \Eprint
  {http://arxiv.org/abs/http://aip.scitation.org/doi/pdf/10.1063/1.1693587}
  {http://aip.scitation.org/doi/pdf/10.1063/1.1693587} \BibitemShut {NoStop}%
\bibitem [{\citenamefont {Gentle}\ and\ \citenamefont
  {Lohr}(1973)}]{Gentle1973-PoF}%
  \BibitemOpen
  \bibfield  {author} {\bibinfo {author} {\bibfnamefont {K.~W.}\ \bibnamefont
  {Gentle}}\ and\ \bibinfo {author} {\bibfnamefont {J.}~\bibnamefont {Lohr}},\
  }\bibfield  {title} {\enquote {\bibinfo {title} {Experimental determination
  of the nonlinear interaction in a one dimensional beam‐plasma system},}\
  }\href {\doibase 10.1063/1.1694543} {\bibfield  {journal} {\bibinfo
  {journal} {The Physics of Fluids}\ }\textbf {\bibinfo {volume} {16}},\
  \bibinfo {pages} {1464--1471} (\bibinfo {year} {1973})},\ \Eprint
  {http://arxiv.org/abs/http://aip.scitation.org/doi/pdf/10.1063/1.1694543}
  {http://aip.scitation.org/doi/pdf/10.1063/1.1694543} \BibitemShut {NoStop}%
\bibitem [{\citenamefont {Maggs}(1976)}]{Maggs1976-JGR}%
  \BibitemOpen
  \bibfield  {author} {\bibinfo {author} {\bibfnamefont {J.~E.}\ \bibnamefont
  {Maggs}},\ }\bibfield  {title} {\enquote {\bibinfo {title} {coherent
  generation of {VLF} hiss},}\ }\href {\doibase 10.1029/JA081i010p01707}
  {\bibfield  {journal} {\bibinfo  {journal} {Journal of Geophysical Research}\
  }\textbf {\bibinfo {volume} {81}},\ \bibinfo {pages} {1707--1724} (\bibinfo
  {year} {1976})}\BibitemShut {NoStop}%
\bibitem [{\citenamefont {Gary}\ and\ \citenamefont
  {Feldman}(1977)}]{Gary1977-JGR}%
  \BibitemOpen
  \bibfield  {author} {\bibinfo {author} {\bibfnamefont {S.~P.}\ \bibnamefont
  {Gary}}\ and\ \bibinfo {author} {\bibfnamefont {W.~C.}\ \bibnamefont
  {Feldman}},\ }\bibfield  {title} {\enquote {\bibinfo {title} {Solar wind heat
  flux regulation by the whistler instability},}\ }\href {\doibase
  10.1029/JA082i007p01087} {\bibfield  {journal} {\bibinfo  {journal} {Journal
  of Geophysical Research}\ }\textbf {\bibinfo {volume} {82}},\ \bibinfo
  {pages} {1087--1094} (\bibinfo {year} {1977})}\BibitemShut {NoStop}%
\bibitem [{\citenamefont {Huang}\ \emph {et~al.}(2016)\citenamefont {Huang},
  \citenamefont {Fu}, \citenamefont {Yuan}, \citenamefont {Vaivads},
  \citenamefont {Khotyaintsev}, \citenamefont {Retino}, \citenamefont {Zhou},
  \citenamefont {Graham}, \citenamefont {Fujimoto}, \citenamefont {Sahraoui},
  \citenamefont {Deng}, \citenamefont {Ni}, \citenamefont {Pang}, \citenamefont
  {Fu}, \citenamefont {Wang},\ and\ \citenamefont {Zhou}}]{Huang2016-JGR}%
  \BibitemOpen
  \bibfield  {author} {\bibinfo {author} {\bibfnamefont {S.~Y.}\ \bibnamefont
  {Huang}}, \bibinfo {author} {\bibfnamefont {H.~S.}\ \bibnamefont {Fu}},
  \bibinfo {author} {\bibfnamefont {Z.~G.}\ \bibnamefont {Yuan}}, \bibinfo
  {author} {\bibfnamefont {A.}~\bibnamefont {Vaivads}}, \bibinfo {author}
  {\bibfnamefont {Y.~V.}\ \bibnamefont {Khotyaintsev}}, \bibinfo {author}
  {\bibfnamefont {A.}~\bibnamefont {Retino}}, \bibinfo {author} {\bibfnamefont
  {M.}~\bibnamefont {Zhou}}, \bibinfo {author} {\bibfnamefont {D.~B.}\
  \bibnamefont {Graham}}, \bibinfo {author} {\bibfnamefont {K.}~\bibnamefont
  {Fujimoto}}, \bibinfo {author} {\bibfnamefont {F.}~\bibnamefont {Sahraoui}},
  \bibinfo {author} {\bibfnamefont {X.~H.}\ \bibnamefont {Deng}}, \bibinfo
  {author} {\bibfnamefont {B.}~\bibnamefont {Ni}}, \bibinfo {author}
  {\bibfnamefont {Y.}~\bibnamefont {Pang}}, \bibinfo {author} {\bibfnamefont
  {S.}~\bibnamefont {Fu}}, \bibinfo {author} {\bibfnamefont {D.~D.}\
  \bibnamefont {Wang}}, \ and\ \bibinfo {author} {\bibfnamefont
  {X.}~\bibnamefont {Zhou}},\ }\bibfield  {title} {\enquote {\bibinfo {title}
  {Two types of whistler waves in the hall reconnection region},}\ }\href
  {\doibase 10.1002/2016JA022650} {\bibfield  {journal} {\bibinfo  {journal}
  {Journal of Geophysical Research: Space Physics}\ }\textbf {\bibinfo {volume}
  {121}},\ \bibinfo {pages} {6639--6646} (\bibinfo {year} {2016})},\ \bibinfo
  {note} {2016JA022650}\BibitemShut {NoStop}%
\bibitem [{\citenamefont {Newman}\ \emph {et~al.}(2001)\citenamefont {Newman},
  \citenamefont {Goldman}, \citenamefont {Ergun},\ and\ \citenamefont
  {Mangeney}}]{Newman2001-PRL}%
  \BibitemOpen
  \bibfield  {author} {\bibinfo {author} {\bibfnamefont {D.~L.}\ \bibnamefont
  {Newman}}, \bibinfo {author} {\bibfnamefont {M.~V.}\ \bibnamefont {Goldman}},
  \bibinfo {author} {\bibfnamefont {R.~E.}\ \bibnamefont {Ergun}}, \ and\
  \bibinfo {author} {\bibfnamefont {A.}~\bibnamefont {Mangeney}},\ }\bibfield
  {title} {\enquote {\bibinfo {title} {Formation of double layers and electron
  holes in a current-driven space plasma},}\ }\href {\doibase
  10.1103/PhysRevLett.87.255001} {\bibfield  {journal} {\bibinfo  {journal}
  {Phys. Rev. Lett.}\ }\textbf {\bibinfo {volume} {87}},\ \bibinfo {pages}
  {255001} (\bibinfo {year} {2001})}\BibitemShut {NoStop}%
\bibitem [{\citenamefont {Winckler}(1980)}]{Winckler1980-ROG}%
  \BibitemOpen
  \bibfield  {author} {\bibinfo {author} {\bibfnamefont {J.~R.}\ \bibnamefont
  {Winckler}},\ }\bibfield  {title} {\enquote {\bibinfo {title} {The
  application of artificial electron beams to magnetospheric research},}\
  }\href {\doibase 10.1029/RG018i003p00659} {\bibfield  {journal} {\bibinfo
  {journal} {Reviews of Geophysics}\ }\textbf {\bibinfo {volume} {18}},\
  \bibinfo {pages} {659--682} (\bibinfo {year} {1980})}\BibitemShut {NoStop}%
\bibitem [{\citenamefont {Stenzel}(1977)}]{Stenzel1977-JGR}%
  \BibitemOpen
  \bibfield  {author} {\bibinfo {author} {\bibfnamefont {R.~L.}\ \bibnamefont
  {Stenzel}},\ }\bibfield  {title} {\enquote {\bibinfo {title} {Observation of
  beam-generated {VLF} hiss in a large laboratory plasma},}\ }\href {\doibase
  10.1029/JA082i029p04805} {\bibfield  {journal} {\bibinfo  {journal} {Journal
  of Geophysical Research}\ }\textbf {\bibinfo {volume} {82}},\ \bibinfo
  {pages} {4805--4814} (\bibinfo {year} {1977})}\BibitemShut {NoStop}%
\bibitem [{\citenamefont {Krafft}\ \emph {et~al.}(1994)\citenamefont {Krafft},
  \citenamefont {Th\'evenet}, \citenamefont {Matthieussent}, \citenamefont
  {Lundin}, \citenamefont {Belmont}, \citenamefont {Lemb\`ege}, \citenamefont
  {Solomon}, \citenamefont {Lavergnat},\ and\ \citenamefont
  {Lehner}}]{Krafft1994-PRL}%
  \BibitemOpen
  \bibfield  {author} {\bibinfo {author} {\bibfnamefont {C.}~\bibnamefont
  {Krafft}}, \bibinfo {author} {\bibfnamefont {P.}~\bibnamefont {Th\'evenet}},
  \bibinfo {author} {\bibfnamefont {G.}~\bibnamefont {Matthieussent}}, \bibinfo
  {author} {\bibfnamefont {B.}~\bibnamefont {Lundin}}, \bibinfo {author}
  {\bibfnamefont {G.}~\bibnamefont {Belmont}}, \bibinfo {author} {\bibfnamefont
  {B.}~\bibnamefont {Lemb\`ege}}, \bibinfo {author} {\bibfnamefont
  {J.}~\bibnamefont {Solomon}}, \bibinfo {author} {\bibfnamefont
  {J.}~\bibnamefont {Lavergnat}}, \ and\ \bibinfo {author} {\bibfnamefont
  {T.}~\bibnamefont {Lehner}},\ }\bibfield  {title} {\enquote {\bibinfo {title}
  {Whistler wave emission by a modulated electron beam},}\ }\href {\doibase
  10.1103/PhysRevLett.72.649} {\bibfield  {journal} {\bibinfo  {journal} {Phys.
  Rev. Lett.}\ }\textbf {\bibinfo {volume} {72}},\ \bibinfo {pages} {649--652}
  (\bibinfo {year} {1994})}\BibitemShut {NoStop}%
\bibitem [{\citenamefont {Starodubtsev}\ and\ \citenamefont
  {Krafft}(1999)}]{Staro1999-PRL}%
  \BibitemOpen
  \bibfield  {author} {\bibinfo {author} {\bibfnamefont {M.}~\bibnamefont
  {Starodubtsev}}\ and\ \bibinfo {author} {\bibfnamefont {C.}~\bibnamefont
  {Krafft}},\ }\bibfield  {title} {\enquote {\bibinfo {title} {Resonant
  cyclotron emission of whistler waves by a modulated electron beam},}\ }\href
  {\doibase 10.1103/PhysRevLett.83.1335} {\bibfield  {journal} {\bibinfo
  {journal} {Phys. Rev. Lett.}\ }\textbf {\bibinfo {volume} {83}},\ \bibinfo
  {pages} {1335--1338} (\bibinfo {year} {1999})}\BibitemShut {NoStop}%
\bibitem [{\citenamefont {Morse}\ and\ \citenamefont
  {Nielson}(1969)}]{Morese1969-PoF}%
  \BibitemOpen
  \bibfield  {author} {\bibinfo {author} {\bibfnamefont {R.~L.}\ \bibnamefont
  {Morse}}\ and\ \bibinfo {author} {\bibfnamefont {C.~W.}\ \bibnamefont
  {Nielson}},\ }\bibfield  {title} {\enquote {\bibinfo {title} {Numerical
  simulation of warm two‐beam plasma},}\ }\href {\doibase 10.1063/1.1692361}
  {\bibfield  {journal} {\bibinfo  {journal} {The Physics of Fluids}\ }\textbf
  {\bibinfo {volume} {12}},\ \bibinfo {pages} {2418--2425} (\bibinfo {year}
  {1969})},\ \Eprint
  {http://arxiv.org/abs/http://aip.scitation.org/doi/pdf/10.1063/1.1692361}
  {http://aip.scitation.org/doi/pdf/10.1063/1.1692361} \BibitemShut {NoStop}%
\bibitem [{\citenamefont {Omura}\ and\ \citenamefont
  {Matsumoto}(1987)}]{Omura1987-JGR}%
  \BibitemOpen
  \bibfield  {author} {\bibinfo {author} {\bibfnamefont {Y.}~\bibnamefont
  {Omura}}\ and\ \bibinfo {author} {\bibfnamefont {H.}~\bibnamefont
  {Matsumoto}},\ }\bibfield  {title} {\enquote {\bibinfo {title} {Competing
  processes of whistler and electrostatic instabilities in the
  magnetosphere},}\ }\href {\doibase 10.1029/JA092iA08p08649} {\bibfield
  {journal} {\bibinfo  {journal} {Journal of Geophysical Research: Space
  Physics}\ }\textbf {\bibinfo {volume} {92}},\ \bibinfo {pages} {8649--8659}
  (\bibinfo {year} {1987})}\BibitemShut {NoStop}%
\bibitem [{\citenamefont {Omura}\ and\ \citenamefont
  {Matsumoto}(1988)}]{Omura1988-GRL}%
  \BibitemOpen
  \bibfield  {author} {\bibinfo {author} {\bibfnamefont {Y.}~\bibnamefont
  {Omura}}\ and\ \bibinfo {author} {\bibfnamefont {H.}~\bibnamefont
  {Matsumoto}},\ }\bibfield  {title} {\enquote {\bibinfo {title} {Computer
  experiments on whistler and plasma wave emissions for spacelab-2 electron
  beam},}\ }\href {\doibase 10.1029/GL015i004p00319} {\bibfield  {journal}
  {\bibinfo  {journal} {Geophysical Research Letters}\ }\textbf {\bibinfo
  {volume} {15}},\ \bibinfo {pages} {319--322} (\bibinfo {year}
  {1988})}\BibitemShut {NoStop}%
\bibitem [{\citenamefont {Pritchett}, \citenamefont {Karimabadi},\ and\
  \citenamefont {Omidi}(1989)}]{Pritchett1989-GRL}%
  \BibitemOpen
  \bibfield  {author} {\bibinfo {author} {\bibfnamefont {P.~L.}\ \bibnamefont
  {Pritchett}}, \bibinfo {author} {\bibfnamefont {H.}~\bibnamefont
  {Karimabadi}}, \ and\ \bibinfo {author} {\bibfnamefont {N.}~\bibnamefont
  {Omidi}},\ }\bibfield  {title} {\enquote {\bibinfo {title} {Generation
  mechanism of whistler waves produced by electron beam injection in space},}\
  }\href {\doibase 10.1029/GL016i008p00883} {\bibfield  {journal} {\bibinfo
  {journal} {Geophysical Research Letters}\ }\textbf {\bibinfo {volume} {16}},\
  \bibinfo {pages} {883--886} (\bibinfo {year} {1989})}\BibitemShut {NoStop}%
\bibitem [{\citenamefont {Gary}\ \emph {et~al.}(2000)\citenamefont {Gary},
  \citenamefont {Kazimura}, \citenamefont {Li},\ and\ \citenamefont
  {Sakai}}]{Gary2000-PoP}%
  \BibitemOpen
  \bibfield  {author} {\bibinfo {author} {\bibfnamefont {S.~P.}\ \bibnamefont
  {Gary}}, \bibinfo {author} {\bibfnamefont {Y.}~\bibnamefont {Kazimura}},
  \bibinfo {author} {\bibfnamefont {H.}~\bibnamefont {Li}}, \ and\ \bibinfo
  {author} {\bibfnamefont {J.-I.}\ \bibnamefont {Sakai}},\ }\bibfield  {title}
  {\enquote {\bibinfo {title} {Simulations of electron/electron instabilities:
  Electromagnetic fluctuations},}\ }\href {\doibase 10.1063/1.873829}
  {\bibfield  {journal} {\bibinfo  {journal} {Physics of Plasmas}\ }\textbf
  {\bibinfo {volume} {7}},\ \bibinfo {pages} {448--456} (\bibinfo {year}
  {2000})},\ \Eprint {http://arxiv.org/abs/http://dx.doi.org/10.1063/1.873829}
  {http://dx.doi.org/10.1063/1.873829} \BibitemShut {NoStop}%
\bibitem [{\citenamefont {Fu}\ \emph {et~al.}(2014)\citenamefont {Fu},
  \citenamefont {Cowee}, \citenamefont {Liu}, \citenamefont {Gary},\ and\
  \citenamefont {Winske}}]{Fu2014-PoP}%
  \BibitemOpen
  \bibfield  {author} {\bibinfo {author} {\bibfnamefont {X.~R.}\ \bibnamefont
  {Fu}}, \bibinfo {author} {\bibfnamefont {M.~M.}\ \bibnamefont {Cowee}},
  \bibinfo {author} {\bibfnamefont {K.}~\bibnamefont {Liu}}, \bibinfo {author}
  {\bibfnamefont {S.~P.}\ \bibnamefont {Gary}}, \ and\ \bibinfo {author}
  {\bibfnamefont {D.}~\bibnamefont {Winske}},\ }\bibfield  {title} {\enquote
  {\bibinfo {title} {Particle-in-cell simulations of velocity scattering of an
  anisotropic electron beam by electrostatic and electromagnetic
  instabilities},}\ }\href {\doibase 10.1063/1.4870632} {\bibfield  {journal}
  {\bibinfo  {journal} {Physics of Plasmas}\ }\textbf {\bibinfo {volume}
  {21}},\ \bibinfo {pages} {042108} (\bibinfo {year} {2014})},\ \Eprint
  {http://arxiv.org/abs/http://dx.doi.org/10.1063/1.4870632}
  {http://dx.doi.org/10.1063/1.4870632} \BibitemShut {NoStop}%
\bibitem [{\citenamefont {Che}\ \emph {et~al.}(2017)\citenamefont {Che},
  \citenamefont {Goldstein}, \citenamefont {Diamond},\ and\ \citenamefont
  {Sagdeev}}]{Che2017-PNAS}%
  \BibitemOpen
  \bibfield  {author} {\bibinfo {author} {\bibfnamefont {H.}~\bibnamefont
  {Che}}, \bibinfo {author} {\bibfnamefont {M.~L.}\ \bibnamefont {Goldstein}},
  \bibinfo {author} {\bibfnamefont {P.~H.}\ \bibnamefont {Diamond}}, \ and\
  \bibinfo {author} {\bibfnamefont {R.~Z.}\ \bibnamefont {Sagdeev}},\
  }\bibfield  {title} {\enquote {\bibinfo {title} {How electron two-stream
  instability drives cyclic langmuir collapse and continuous coherent
  emission},}\ }\href {\doibase 10.1073/pnas.1614055114} {\bibfield  {journal}
  {\bibinfo  {journal} {Proceedings of the National Academy of Sciences}\
  }\textbf {\bibinfo {volume} {114}},\ \bibinfo {pages} {1502--1507} (\bibinfo
  {year} {2017})},\ \Eprint
  {http://arxiv.org/abs/http://www.pnas.org/content/114/7/1502.full.pdf}
  {http://www.pnas.org/content/114/7/1502.full.pdf} \BibitemShut {NoStop}%
\bibitem [{\citenamefont {Van~Compernolle}\ \emph {et~al.}(2015)\citenamefont
  {Van~Compernolle}, \citenamefont {An}, \citenamefont {Bortnik}, \citenamefont
  {Thorne}, \citenamefont {Pribyl},\ and\ \citenamefont
  {Gekelman}}]{VanCompernolle2015-PRL}%
  \BibitemOpen
  \bibfield  {author} {\bibinfo {author} {\bibfnamefont {B.}~\bibnamefont
  {Van~Compernolle}}, \bibinfo {author} {\bibfnamefont {X.}~\bibnamefont {An}},
  \bibinfo {author} {\bibfnamefont {J.}~\bibnamefont {Bortnik}}, \bibinfo
  {author} {\bibfnamefont {R.~M.}\ \bibnamefont {Thorne}}, \bibinfo {author}
  {\bibfnamefont {P.}~\bibnamefont {Pribyl}}, \ and\ \bibinfo {author}
  {\bibfnamefont {W.}~\bibnamefont {Gekelman}},\ }\bibfield  {title} {\enquote
  {\bibinfo {title} {Excitation of chirping whistler waves in a laboratory
  plasma},}\ }\href {\doibase 10.1103/PhysRevLett.114.245002} {\bibfield
  {journal} {\bibinfo  {journal} {Phys. Rev. Lett.}\ }\textbf {\bibinfo
  {volume} {114}},\ \bibinfo {pages} {245002} (\bibinfo {year}
  {2015})}\BibitemShut {NoStop}%
\bibitem [{\citenamefont {Van~Compernolle}\ \emph {et~al.}(2016)\citenamefont
  {Van~Compernolle}, \citenamefont {An}, \citenamefont {Bortnik}, \citenamefont
  {Thorne}, \citenamefont {Pribyl},\ and\ \citenamefont
  {Gekelman}}]{VanCompernolle2016-PRLerr}%
  \BibitemOpen
  \bibfield  {author} {\bibinfo {author} {\bibfnamefont {B.}~\bibnamefont
  {Van~Compernolle}}, \bibinfo {author} {\bibfnamefont {X.}~\bibnamefont {An}},
  \bibinfo {author} {\bibfnamefont {J.}~\bibnamefont {Bortnik}}, \bibinfo
  {author} {\bibfnamefont {R.~M.}\ \bibnamefont {Thorne}}, \bibinfo {author}
  {\bibfnamefont {P.}~\bibnamefont {Pribyl}}, \ and\ \bibinfo {author}
  {\bibfnamefont {W.}~\bibnamefont {Gekelman}},\ }\bibfield  {title} {\enquote
  {\bibinfo {title} {Erratum: Excitation of chirping whistler waves in a
  laboratory plasma [phys. rev. lett. 114, 245002 (2015)]},}\ }\href {\doibase
  10.1103/PhysRevLett.117.059901} {\bibfield  {journal} {\bibinfo  {journal}
  {Phys. Rev. Lett.}\ }\textbf {\bibinfo {volume} {117}},\ \bibinfo {pages}
  {059901} (\bibinfo {year} {2016})}\BibitemShut {NoStop}%
\bibitem [{\citenamefont {An}\ \emph {et~al.}(2016)\citenamefont {An},
  \citenamefont {Van~Compernolle}, \citenamefont {Bortnik}, \citenamefont
  {Thorne}, \citenamefont {Chen},\ and\ \citenamefont {Li}}]{An2016-GRL}%
  \BibitemOpen
  \bibfield  {author} {\bibinfo {author} {\bibfnamefont {X.}~\bibnamefont
  {An}}, \bibinfo {author} {\bibfnamefont {B.}~\bibnamefont {Van~Compernolle}},
  \bibinfo {author} {\bibfnamefont {J.}~\bibnamefont {Bortnik}}, \bibinfo
  {author} {\bibfnamefont {R.~M.}\ \bibnamefont {Thorne}}, \bibinfo {author}
  {\bibfnamefont {L.}~\bibnamefont {Chen}}, \ and\ \bibinfo {author}
  {\bibfnamefont {W.}~\bibnamefont {Li}},\ }\bibfield  {title} {\enquote
  {\bibinfo {title} {Resonant excitation of whistler waves by a helical
  electron beam},}\ }\href {\doibase 10.1002/2015GL067126} {\bibfield
  {journal} {\bibinfo  {journal} {Geophysical Research Letters}\ }\textbf
  {\bibinfo {volume} {43}},\ \bibinfo {pages} {2413--2421} (\bibinfo {year}
  {2016})},\ \bibinfo {note} {2015GL067126}\BibitemShut {NoStop}%
\bibitem [{\citenamefont {Compernolle}\ \emph {et~al.}(2017)\citenamefont
  {Compernolle}, \citenamefont {An}, \citenamefont {Bortnik}, \citenamefont
  {Thorne}, \citenamefont {Pribyl},\ and\ \citenamefont
  {Gekelman}}]{VanCompernolle2017-PPCF}%
  \BibitemOpen
  \bibfield  {author} {\bibinfo {author} {\bibfnamefont {B.~V.}\ \bibnamefont
  {Compernolle}}, \bibinfo {author} {\bibfnamefont {X.}~\bibnamefont {An}},
  \bibinfo {author} {\bibfnamefont {J.}~\bibnamefont {Bortnik}}, \bibinfo
  {author} {\bibfnamefont {R.~M.}\ \bibnamefont {Thorne}}, \bibinfo {author}
  {\bibfnamefont {P.}~\bibnamefont {Pribyl}}, \ and\ \bibinfo {author}
  {\bibfnamefont {W.}~\bibnamefont {Gekelman}},\ }\bibfield  {title} {\enquote
  {\bibinfo {title} {Laboratory simulation of magnetospheric chorus wave
  generation},}\ }\href {http://stacks.iop.org/0741-3335/59/i=1/a=014016}
  {\bibfield  {journal} {\bibinfo  {journal} {Plasma Physics and Controlled
  Fusion}\ }\textbf {\bibinfo {volume} {59}},\ \bibinfo {pages} {014016}
  (\bibinfo {year} {2017})}\BibitemShut {NoStop}%
\bibitem [{\citenamefont {Gekelman}\ \emph {et~al.}(2016)\citenamefont
  {Gekelman}, \citenamefont {Pribyl}, \citenamefont {Lucky}, \citenamefont
  {Drandell}, \citenamefont {Leneman}, \citenamefont {Maggs}, \citenamefont
  {Vincena}, \citenamefont {Compernolle}, \citenamefont {Tripathi},
  \citenamefont {Morales}, \citenamefont {Carter}, \citenamefont {Wang},\ and\
  \citenamefont {DeHaas}}]{Gekelman2016-RSI}%
  \BibitemOpen
  \bibfield  {author} {\bibinfo {author} {\bibfnamefont {W.}~\bibnamefont
  {Gekelman}}, \bibinfo {author} {\bibfnamefont {P.}~\bibnamefont {Pribyl}},
  \bibinfo {author} {\bibfnamefont {Z.}~\bibnamefont {Lucky}}, \bibinfo
  {author} {\bibfnamefont {M.}~\bibnamefont {Drandell}}, \bibinfo {author}
  {\bibfnamefont {D.}~\bibnamefont {Leneman}}, \bibinfo {author} {\bibfnamefont
  {J.}~\bibnamefont {Maggs}}, \bibinfo {author} {\bibfnamefont
  {S.}~\bibnamefont {Vincena}}, \bibinfo {author} {\bibfnamefont {B.~V.}\
  \bibnamefont {Compernolle}}, \bibinfo {author} {\bibfnamefont {S.~K.~P.}\
  \bibnamefont {Tripathi}}, \bibinfo {author} {\bibfnamefont {G.}~\bibnamefont
  {Morales}}, \bibinfo {author} {\bibfnamefont {T.~A.}\ \bibnamefont {Carter}},
  \bibinfo {author} {\bibfnamefont {Y.}~\bibnamefont {Wang}}, \ and\ \bibinfo
  {author} {\bibfnamefont {T.}~\bibnamefont {DeHaas}},\ }\bibfield  {title}
  {\enquote {\bibinfo {title} {The upgraded large plasma device, a machine for
  studying frontier basic plasma physics},}\ }\href {\doibase
  10.1063/1.4941079} {\bibfield  {journal} {\bibinfo  {journal} {Review of
  Scientific Instruments}\ }\textbf {\bibinfo {volume} {87}},\ \bibinfo {pages}
  {025105} (\bibinfo {year} {2016})},\ \Eprint
  {http://arxiv.org/abs/http://dx.doi.org/10.1063/1.4941079}
  {http://dx.doi.org/10.1063/1.4941079} \BibitemShut {NoStop}%
\bibitem [{\citenamefont {Decyk}(2007)}]{Decyk2007-CPC}%
  \BibitemOpen
  \bibfield  {author} {\bibinfo {author} {\bibfnamefont {V.~K.}\ \bibnamefont
  {Decyk}},\ }\bibfield  {title} {\enquote {\bibinfo {title} {{UPIC}: {A}
  framework for massively parallel particle-in-cell codes},}\ }\href {\doibase
  http://dx.doi.org/10.1016/j.cpc.2007.02.092} {\bibfield  {journal} {\bibinfo
  {journal} {Computer Physics Communications}\ }\textbf {\bibinfo {volume}
  {177}},\ \bibinfo {pages} {95 -- 97} (\bibinfo {year} {2007})},\ \bibinfo
  {note} {proceedings of the Conference on Computational Physics
  2006}\BibitemShut {NoStop}%
\bibitem [{PIC()}]{PICKSC-skeleton}%
  \BibitemOpen
  \href@noop {} {}\bibinfo {howpublished}
  {\url{http://picksc.idre.ucla.edu/software/skeleton-code/}}\BibitemShut
  {NoStop}%
\bibitem [{\citenamefont {Hughes}\ \emph {et~al.}(2016)\citenamefont {Hughes},
  \citenamefont {Wang}, \citenamefont {Decyk},\ and\ \citenamefont
  {Gary}}]{Hughes2016-PoP}%
  \BibitemOpen
  \bibfield  {author} {\bibinfo {author} {\bibfnamefont {R.~S.}\ \bibnamefont
  {Hughes}}, \bibinfo {author} {\bibfnamefont {J.}~\bibnamefont {Wang}},
  \bibinfo {author} {\bibfnamefont {V.~K.}\ \bibnamefont {Decyk}}, \ and\
  \bibinfo {author} {\bibfnamefont {S.~P.}\ \bibnamefont {Gary}},\ }\bibfield
  {title} {\enquote {\bibinfo {title} {Effects of variations in electron
  thermal velocity on the whistler anisotropy instability: Particle-in-cell
  simulations},}\ }\href {\doibase 10.1063/1.4945748} {\bibfield  {journal}
  {\bibinfo  {journal} {Physics of Plasmas}\ }\textbf {\bibinfo {volume}
  {23}},\ \bibinfo {pages} {042106} (\bibinfo {year} {2016})},\ \Eprint
  {http://arxiv.org/abs/http://dx.doi.org/10.1063/1.4945748}
  {http://dx.doi.org/10.1063/1.4945748} \BibitemShut {NoStop}%
\bibitem [{\citenamefont {Schriver}\ \emph {et~al.}(2010)\citenamefont
  {Schriver}, \citenamefont {Ashour-Abdalla}, \citenamefont {Coroniti},
  \citenamefont {LeBoeuf}, \citenamefont {Decyk}, \citenamefont {Travnicek},
  \citenamefont {Santolík}, \citenamefont {Winningham}, \citenamefont
  {Pickett}, \citenamefont {Goldstein},\ and\ \citenamefont
  {Fazakerley}}]{Schriver2010-JGR}%
  \BibitemOpen
  \bibfield  {author} {\bibinfo {author} {\bibfnamefont {D.}~\bibnamefont
  {Schriver}}, \bibinfo {author} {\bibfnamefont {M.}~\bibnamefont
  {Ashour-Abdalla}}, \bibinfo {author} {\bibfnamefont {F.~V.}\ \bibnamefont
  {Coroniti}}, \bibinfo {author} {\bibfnamefont {J.~N.}\ \bibnamefont
  {LeBoeuf}}, \bibinfo {author} {\bibfnamefont {V.}~\bibnamefont {Decyk}},
  \bibinfo {author} {\bibfnamefont {P.}~\bibnamefont {Travnicek}}, \bibinfo
  {author} {\bibfnamefont {O.}~\bibnamefont {Santolík}}, \bibinfo {author}
  {\bibfnamefont {D.}~\bibnamefont {Winningham}}, \bibinfo {author}
  {\bibfnamefont {J.~S.}\ \bibnamefont {Pickett}}, \bibinfo {author}
  {\bibfnamefont {M.~L.}\ \bibnamefont {Goldstein}}, \ and\ \bibinfo {author}
  {\bibfnamefont {A.~N.}\ \bibnamefont {Fazakerley}},\ }\bibfield  {title}
  {\enquote {\bibinfo {title} {Generation of whistler mode emissions in the
  inner magnetosphere: An event study},}\ }\href {\doibase
  10.1029/2009JA014932} {\bibfield  {journal} {\bibinfo  {journal} {Journal of
  Geophysical Research: Space Physics}\ }\textbf {\bibinfo {volume} {115}},\
  \bibinfo {pages} {{A}00{F}17} (\bibinfo {year} {2010})}\BibitemShut {NoStop}%
\bibitem [{\citenamefont {An}\ \emph {et~al.}(2017)\citenamefont {An},
  \citenamefont {Yue}, \citenamefont {Bortnik}, \citenamefont {Decyk},
  \citenamefont {Li},\ and\ \citenamefont {Thorne}}]{An2017-JGR}%
  \BibitemOpen
  \bibfield  {author} {\bibinfo {author} {\bibfnamefont {X.}~\bibnamefont
  {An}}, \bibinfo {author} {\bibfnamefont {C.}~\bibnamefont {Yue}}, \bibinfo
  {author} {\bibfnamefont {J.}~\bibnamefont {Bortnik}}, \bibinfo {author}
  {\bibfnamefont {V.}~\bibnamefont {Decyk}}, \bibinfo {author} {\bibfnamefont
  {W.}~\bibnamefont {Li}}, \ and\ \bibinfo {author} {\bibfnamefont {R.~M.}\
  \bibnamefont {Thorne}},\ }\bibfield  {title} {\enquote {\bibinfo {title} {On
  the parameter dependence of the whistler anisotropy instability},}\ }\href
  {\doibase 10.1002/2017JA023895} {\bibfield  {journal} {\bibinfo  {journal}
  {Journal of Geophysical Research: Space Physics}\ }\textbf {\bibinfo {volume}
  {122}},\ \bibinfo {pages} {2001--2009} (\bibinfo {year} {2017})},\ \bibinfo
  {note} {2017JA023895}\BibitemShut {NoStop}%
\bibitem [{\citenamefont {Busnardo-Neto}\ \emph {et~al.}(1977)\citenamefont
  {Busnardo-Neto}, \citenamefont {Pritchett}, \citenamefont {Lin},\ and\
  \citenamefont {Dawson}}]{BUSNARDONETO1977300}%
  \BibitemOpen
  \bibfield  {author} {\bibinfo {author} {\bibfnamefont {J.}~\bibnamefont
  {Busnardo-Neto}}, \bibinfo {author} {\bibfnamefont {P.}~\bibnamefont
  {Pritchett}}, \bibinfo {author} {\bibfnamefont {A.}~\bibnamefont {Lin}}, \
  and\ \bibinfo {author} {\bibfnamefont {J.}~\bibnamefont {Dawson}},\
  }\bibfield  {title} {\enquote {\bibinfo {title} {A self-consistent
  magnetostatic particle code for numerical simulation of plasmas},}\ }\href
  {\doibase http://dx.doi.org/10.1016/0021-9991(77)90096-1} {\bibfield
  {journal} {\bibinfo  {journal} {Journal of Computational Physics}\ }\textbf
  {\bibinfo {volume} {23}},\ \bibinfo {pages} {300 -- 312} (\bibinfo {year}
  {1977})}\BibitemShut {NoStop}%
\bibitem [{\citenamefont {Geary}\ \emph {et~al.}(1986)\citenamefont {Geary},
  \citenamefont {Tajima}, \citenamefont {Leboeuf}, \citenamefont {Zaidman},\
  and\ \citenamefont {Han}}]{GEARY1986313}%
  \BibitemOpen
  \bibfield  {author} {\bibinfo {author} {\bibfnamefont {J.}~\bibnamefont
  {Geary}}, \bibinfo {author} {\bibfnamefont {T.}~\bibnamefont {Tajima}},
  \bibinfo {author} {\bibfnamefont {J.-N.}\ \bibnamefont {Leboeuf}}, \bibinfo
  {author} {\bibfnamefont {E.}~\bibnamefont {Zaidman}}, \ and\ \bibinfo
  {author} {\bibfnamefont {J.}~\bibnamefont {Han}},\ }\bibfield  {title}
  {\enquote {\bibinfo {title} {Two- and three-dimensional magnetoinductive
  particle codes with guiding center electron motion},}\ }\href {\doibase
  http://dx.doi.org/10.1016/0010-4655(86)90002-0} {\bibfield  {journal}
  {\bibinfo  {journal} {Computer Physics Communications}\ }\textbf {\bibinfo
  {volume} {42}},\ \bibinfo {pages} {313 -- 331} (\bibinfo {year}
  {1986})}\BibitemShut {NoStop}%
\bibitem [{\citenamefont {Hewett}(1985)}]{Hewett1985}%
  \BibitemOpen
  \bibfield  {author} {\bibinfo {author} {\bibfnamefont {D.~W.}\ \bibnamefont
  {Hewett}},\ }\enquote {\bibinfo {title} {Elimination of electromagnetic
  radiation in plasma simulation: The darwin or magneto inductive
  approximation},}\ in\ \href {\doibase 10.1007/978-94-009-5454-0\_3} {\emph
  {\bibinfo {booktitle} {Space Plasma Simulations: Proceedings of the Second
  International School for Space Simulations, Kapaa, Hawaii, February 4--15,
  1985}}},\ \bibinfo {editor} {edited by\ \bibinfo {editor} {\bibfnamefont
  {M.}~\bibnamefont {Ashour-Abdalla}}\ and\ \bibinfo {editor} {\bibfnamefont
  {D.~A.}\ \bibnamefont {Dutton}}}\ (\bibinfo  {publisher} {Springer
  Netherlands},\ \bibinfo {address} {Dordrecht},\ \bibinfo {year} {1985})\ pp.\
  \bibinfo {pages} {29--40}\BibitemShut {NoStop}%
\bibitem [{\citenamefont {Grossmann}\ and\ \citenamefont
  {Morlet}(1984)}]{Grossmann1984-JMA}%
  \BibitemOpen
  \bibfield  {author} {\bibinfo {author} {\bibfnamefont {A.}~\bibnamefont
  {Grossmann}}\ and\ \bibinfo {author} {\bibfnamefont {J.}~\bibnamefont
  {Morlet}},\ }\bibfield  {title} {\enquote {\bibinfo {title} {Decomposition of
  hardy functions into square integrable wavelets of constant shape},}\ }\href
  {\doibase 10.1137/0515056} {\bibfield  {journal} {\bibinfo  {journal} {SIAM
  Journal on Mathematical Analysis}\ }\textbf {\bibinfo {volume} {15}},\
  \bibinfo {pages} {723--736} (\bibinfo {year} {1984})},\ \Eprint
  {http://arxiv.org/abs/http://dx.doi.org/10.1137/0515056}
  {http://dx.doi.org/10.1137/0515056} \BibitemShut {NoStop}%
\bibitem [{\citenamefont {Goupillaud}, \citenamefont {Grossmann},\ and\
  \citenamefont {Morlet}(1984)}]{Goupillaud1984-Geoeploration}%
  \BibitemOpen
  \bibfield  {author} {\bibinfo {author} {\bibfnamefont {P.}~\bibnamefont
  {Goupillaud}}, \bibinfo {author} {\bibfnamefont {A.}~\bibnamefont
  {Grossmann}}, \ and\ \bibinfo {author} {\bibfnamefont {J.}~\bibnamefont
  {Morlet}},\ }\bibfield  {title} {\enquote {\bibinfo {title} {Cycle-octave and
  related transforms in seismic signal analysis},}\ }\href {\doibase
  http://dx.doi.org/10.1016/0016-7142(84)90025-5} {\bibfield  {journal}
  {\bibinfo  {journal} {Geoexploration}\ }\textbf {\bibinfo {volume} {23}},\
  \bibinfo {pages} {85 -- 102} (\bibinfo {year} {1984})}\BibitemShut {NoStop}%
\bibitem [{\citenamefont {Bell}\ and\ \citenamefont
  {Buneman}(1964)}]{Bell1964-PR}%
  \BibitemOpen
  \bibfield  {author} {\bibinfo {author} {\bibfnamefont {T.~F.}\ \bibnamefont
  {Bell}}\ and\ \bibinfo {author} {\bibfnamefont {O.}~\bibnamefont {Buneman}},\
  }\bibfield  {title} {\enquote {\bibinfo {title} {Plasma instability in the
  whistler mode caused by a gyrating electron stream},}\ }\href {\doibase
  10.1103/PhysRev.133.A1300} {\bibfield  {journal} {\bibinfo  {journal} {Phys.
  Rev.}\ }\textbf {\bibinfo {volume} {133}},\ \bibinfo {pages} {A1300--A1302}
  (\bibinfo {year} {1964})}\BibitemShut {NoStop}%
\bibitem [{\citenamefont {Starodubtsev}\ \emph {et~al.}(1999)\citenamefont
  {Starodubtsev}, \citenamefont {Krafft}, \citenamefont {Lundin},\ and\
  \citenamefont {Thévenet}}]{Starodubtsev1999-PoP}%
  \BibitemOpen
  \bibfield  {author} {\bibinfo {author} {\bibfnamefont {M.}~\bibnamefont
  {Starodubtsev}}, \bibinfo {author} {\bibfnamefont {C.}~\bibnamefont
  {Krafft}}, \bibinfo {author} {\bibfnamefont {B.}~\bibnamefont {Lundin}}, \
  and\ \bibinfo {author} {\bibfnamefont {P.}~\bibnamefont {Thévenet}},\
  }\bibfield  {title} {\enquote {\bibinfo {title} {Resonant cherenkov emission
  of whistlers by a modulated electron beam},}\ }\href {\doibase
  10.1063/1.873244} {\bibfield  {journal} {\bibinfo  {journal} {Physics of
  Plasmas}\ }\textbf {\bibinfo {volume} {6}},\ \bibinfo {pages} {2862--2869}
  (\bibinfo {year} {1999})},\ \Eprint
  {http://arxiv.org/abs/http://dx.doi.org/10.1063/1.873244}
  {http://dx.doi.org/10.1063/1.873244} \BibitemShut {NoStop}%
\end{thebibliography}%

\end{document}